\documentclass[11pt,a4paper]{article}
\usepackage{jheppub}

\RequirePackage{lineno}

\usepackage{atlasphysics} 
\graphicspath{{/}}

\usepackage{subfigure}
\usepackage{mathrsfs}
\usepackage{graphicx}
\usepackage{multirow}
\usepackage{rotating}
\usepackage{xspace}
\usepackage{slashed}
\usepackage{amsmath}
\usepackage{placeins}
\usepackage{preprintcover} 

\newcommand{\lumi}{20.3} %GRL for 2011 LP
\newcommand{\alpgen}{{\sc Alpgen}\xspace}
\newcommand{\herwig}{{\sc Herwig}\xspace}
\newcommand{\herwigpp}{{\sc Herwig++}\xspace}
\newcommand{\mcatnlo}{{\sc MC@NLO}}

\newcommand{\sherpa}{{\sc Sherpa}\xspace}
\newcommand{\blackmax}{{\sc Blackmax}\xspace}
\newcommand{\charybdis}{{\sc Charybdis}\xspace}
\newcommand{\geant}{{\sc GEANT4}\xspace}
\newcommand{\sumpt}{\ensuremath{\sum \pt}\xspace}
\newcommand{\MD}{\ensuremath{M_{\mathrm{D}}}\xspace}
\newcommand{\MTH}{\ensuremath{M_{\mathrm{th}}}\xspace}
\newcommand{\MS}{\ensuremath{M_\mathrm{S}}\xspace}

\newcommand{\ratioDescription}{The lower panels show the ratio of the data to the expected background, with the statistical uncertainty on data (points), and separately, the fractional total uncertainty on the background (shaded band).
}

\PreprintCoverPaperTitle{Search for microscopic black holes and string balls in final states with leptons and jets with the ATLAS detector at $\sqrt{s}=8$ TeV}  % paper title

\PreprintIdNumber{CERN-PH-EP-2014-072}  % CERN preprint number

\PreprintCoverAbstract{A search for an excess of events with multiple high transverse momentum objects including charged leptons and jets is presented, using 20.3~fb$^{-1}$ of proton-proton collision data recorded by the ATLAS detector at the Large Hadron Collider in 2012 at a centre-of-mass energy of $\sqrt{s} = 8$~TeV. No excess of events beyond Standard Model expectations is observed.
Using extra-dimensional models for black hole and string ball production and decay, exclusion contours are determined as a function of the mass threshold for production and the fundamental gravity scale for two, four and six extra dimensions. For six extra dimensions, mass thresholds of $4.8$--$6.2$~TeV are excluded at 95\% confidence level, depending on the fundamental gravity scale and model assumptions. Upper limits on the fiducial cross-sections for non-Standard Model production of these final states are set.}  % paper abstract

\PreprintJournalName{JHEP}  % journal name

\title{Search for microscopic black holes and string balls in final states with leptons and jets with the ATLAS detector at $\sqrt{s}=8$ TeV}
\author{The ATLAS Collaboration}
\abstract{
A search for an excess of events with multiple high transverse momentum objects including charged leptons and jets is presented, using 20.3~fb$^{-1}$ of proton-proton collision data recorded by the ATLAS detector at the Large Hadron Collider in 2012 at a centre-of-mass energy of $\sqrt{s} = 8$~TeV. No excess of events beyond Standard Model expectations is observed.
Using extra-dimensional models for black hole and string ball production and decay, exclusion contours are determined as a function of the mass threshold for production and the fundamental gravity scale for two, four and six extra dimensions. For six extra dimensions, mass thresholds of $4.8$--$6.2$~TeV are excluded at 95\% confidence level, depending on the fundamental gravity scale and model assumptions. Upper limits on the fiducial cross-sections for non-Standard Model production of these final states are set.

}

\begin{document}

\maketitle

%%%%%%%%%%%%%%%%%%%%%%%%%%%%%%%%%%%%%%%%%%%%%%%%%%%%%%%%%%%%%%%%%%%%%%%%%%%%%%
\section{Introduction}
\label{sec:introduction}
A long-standing problem in particle physics is the very large difference between two apparently fundamental energy scales: the electroweak scale at ${\cal O}(0.1$~TeV) and the gravitational (Planck) scale  $M_\mathrm{Pl}={\cal O}(10^{16}$~TeV). Models postulating extra spatial dimensions into which the gravitational field propagates attempt to address this hierarchy problem~\cite{Antoniadis:1990ew, ArkaniHamed:1998rs,Antoniadis:1998ig, Randall:1999ee}. In most of these models, the Standard Model (SM) fields are constrained to the one time and three spatial dimensions of our universe, whilst the gravitons also propagate into the $n$ ``bulk'' extra dimensions.  In these models, the fundamental gravitational scale in the full $(n+4)$ space-time dimensions, \MD, is dramatically lower than $M_\mathrm{Pl}$, and represents an {\em effective} scale appropriate for probes of the gravitational interactions at low energies. 
A value of \MD in the TeV range would allow for the production of strong gravitational states such as microscopic black holes at energies accessible at the Large Hadron Collider (LHC)~\cite{LHCProposalPaper, Giddings:2001bu, DimopoulosLandsbergBHatLHC}. Two well-motivated extra-dimensional models are those with large flat extra dimensions (ADD models~\cite{ArkaniHamed:1998rs,Antoniadis:1998ig}) and those with small, usually warped, extra dimensions (RS models~\cite{Randall:1999ee}). This analysis considers ADD models, for which the $n=1$ case is ruled out and the $n = 2$ case is disfavoured by current astrophysical and tabletop experiments~\cite{pdg2012}. Thus, benchmark models with two, four and six additional spatial dimensions are considered.

Estimates of the black hole production cross-sections invoke semiclassical approximations, the validity of which require the production centre-of-mass energy to be significantly above \MD. This motivates the introduction of a production mass threshold $\MTH$, well above \MD. 
In the black hole formation stage, some energy is expected to be lost to gravitational or SM radiation.
This has recently been calculated using numerical simulations of general relativity~\cite{Sperhake:2011xk}.

Once a black hole has formed and settled into a Schwarzschild~\cite{Schwarzschild:1916uq} (non-rotating) or Myers--Perry~\cite{Myers:1986un} (rotating) state, it is assumed to lose mass and angular momentum through the emission of Hawking radiation~\cite{Hawking75}. All types of SM particles are emitted, although the graviton emission spectra have been calculated only for the non-rotating case~\cite{Emparan:2000rs,Cardoso:2005mh}. The emission energy spectrum is characterised by the Hawking temperature, which depends on $n$, and is larger for lower mass and for more rapidly rotating black holes. It is not a pure black-body spectrum, being modified by gravitational transmission coefficients (``grey-body factors")~\cite{greyBodyFactors, Ida:2005ax, Ida:2006tf, Duffy:2005ns, Casals:2005sa, Casals:2006xp}. These encode the probability of transmission through the gravitational field of the black hole, and act primarily to disfavour low-energy emissions. 
The relative particle emissivities depend on $n$, the black hole angular momentum and temperature, and the spin of the emitted particle. 
In the rotating case, the fluxes for vector field emission are enhanced several-fold, due to the effect of super-radiance~\cite{Ida:2006tf,Casals:2006xp}. 
Emissions reducing the angular momentum of the black hole are favoured kinematically. As the black hole evolves, its mass decreases, and, upon approaching \MD, quantum gravitational effects become important and evaporation by emission of Hawking radiation is no longer a suitable model. This is the ``remnant phase", in which the theoretical modelling uncertainties are large. The conventional treatment by the event generators used in LHC simulations is to decay the black hole remnant to a small number of SM particles~\cite{stringBallsDimopoulos2002}.

Strong gravitational states include, in the context of weakly coupled string theory, highly excited string states (string balls)~\cite{GingrichStringStates2008}.\footnote{Hereinafter all references to black holes also apply to string balls, unless otherwise stated.} In these models, the string scale \MS~\cite{Damour:1999aw} and string coupling $g_\mathrm{S}$ define $\MD = g_\mathrm{S}^{-2/(n+2)} \MS$ and determine the string ball properties. Black hole production and evaporation proceeds as described above, except that black holes evolve into highly excited string states once their mass drops below the correspondence point of $\sim \MS / g_\mathrm{S}^{2}$. Thereafter, the string states continue to emit radiation, with a modified characteristic temperature.

The experimental signature of black hole decays is an ensemble of high-energy particles, the composition of which varies both with model assumptions and \MD; for example, a rotating state leads to fewer emissions of more highly energetic particles. However, the universality of the gravitational coupling implies that particles are produced primarily according to the SM degrees of freedom (modified by the relative emissivities). This leads to a branching fraction to final states with at least one charged lepton\footnote{Throughout this paper, ``lepton" denotes electrons and muons only.} of $\sim 15$--$50\%$, where the range is primarily a consequence of varying average multiplicities of the decay for different models and values of the parameters \MD and \MTH. The most significant uncertainties in the theoretical modelling of these states, which motivate exploration through benchmark models, arise from possible losses of mass-energy and angular momentum in the production phase, the lack of a description of graviton emission in the rotating case, and the treatment of the black hole remnant state at masses near \MD. The latter can strongly impact the multiplicity of particles from black hole decays.

This paper describes a search for an excess of events over SM expectations in \lumi~fb$^{-1}$ of ATLAS $pp$ collision data collected at $\sqrt{s}=8$~TeV in 2012. The analysis considers events at high $\sumpt$, defined as the scalar sum of the $\pt$ of the selected reconstructed objects (hadronic jets and leptons), containing at least three high-$\pt$ objects (leptons or jets), at least one of which must be a lepton. It is similar to a previous search~\cite{Aad2012122}, using $\sqrt{s}=7$~TeV data, which excluded at 95\% confidence level (CL) black holes with \MTH$<4.5$ TeV for \MD$ = 1.5$~TeV and $n=6$. Greater sensitivity in this analysis comes from the higher centre-of-mass energy, more integrated luminosity, as well as from the use of fits to improve background estimates at very high values of $\sumpt$. Searches for black holes have also been performed at $\sqrt{s}=8$~TeV in like-sign dimuonic final states~\cite{Aad:2013bw}, as well as predominantly multi-jet final states~\cite{Khachatryan:2013wx}. The limits set by these two analyses, at 95\% CL, for rotating black holes with \MD$ = 1.5$~TeV and $n=6$ are $\MTH > 5.5$~TeV and $\MTH > 6.2$~TeV, respectively. Corresponding limits for $\MD = 4$~TeV and $n=6$ are $\MTH > 4.5$~TeV and $\MTH > 5.6$~TeV. Two-body final states have also been reported elsewhere~\cite{Khachatryan:2013wx, CMSQBH,tagkey2014562,Aad:2013gma}, with sensitivity to so-called quantum black holes, where the mass is close to \MD.

\section{The ATLAS detector}

ATLAS~\cite{Aad:2008zzm} is a multipurpose detector with a forward-backward symmetric cylindrical
geometry and nearly 4$\pi$ coverage in solid angle.\footnote{ATLAS uses a right-handed coordinate system with its origin at the nominal interaction point (IP) in the centre of the detector and the $z$-axis along the beam pipe. The $x$-axis points from the IP to the centre of the LHC ring, and the $y$-axis points upward. Cylindrical coordinates $(r,\phi)$ are used in the transverse plane, $\phi$ being the azimuthal angle around the beam pipe. The pseudorapidity is defined in terms of the polar angle $\theta$ as $\eta=-\ln\tan(\theta/2)$.} Closest to the beamline, the inner detector (ID) utilises fine-granularity pixel and microstrip detectors designed to provide precise track impact parameter and secondary vertex measurements.  These silicon-based detectors cover the pseudorapidity range $|\eta|< 2.5$.  A gas-filled straw-tube tracker complements the silicon tracker at larger radii.  The tracking detectors are immersed in a $2$~T magnetic field produced by a thin superconducting solenoid located in the same cryostat as the barrel electromagnetic (EM) calorimeter.  The EM calorimeters employ lead absorbers and use liquid argon as the active medium.  The barrel EM calorimeter covers $|\eta|<1.5$ and the end-cap EM calorimeters cover $1.4 < |\eta| <3.2$.  Hadronic calorimetry in the region $|\eta|< 1.7$ is performed using steel absorbers and scintillator tiles as the active medium.  Liquid-argon calorimetry with copper absorbers is used in the hadronic end-cap calorimeters, which cover the region $1.5 < |\eta| < 3.2$. The forward calorimeter ($3.1 < |\eta| < 4.9$) uses copper and tungsten as absorber with liquid argon as active material. The muon spectrometer (MS) measures the deflection of muon tracks within $|\eta|< 2.7$, using three stations of precision drift tubes (with cathode strip chambers in the innermost station for $|\eta| > 2.0$)  located in a toroidal magnetic field of approximately $0.5$~T and $1$~T in the central and end-cap regions of ATLAS, respectively. The muon spectrometer is also instrumented with separate trigger chambers covering $|\eta|< 2.4$.  A three-level trigger is used by the ATLAS detector.  The first-level trigger is implemented in custom electronics, using a subset of detector information to reduce the event rate to a design value of $75$~kHz.  The second and third levels use software algorithms to yield a recorded event rate of about $400$~Hz.

\section{Trigger and data selection}
\label{sec:trigger}

The data used in this analysis were recorded in 2012, while the LHC was operating at a centre-of-mass energy of $8$~\TeV. The integrated luminosity is \lumi~fb$^{-1}$. The uncertainty on the integrated luminosity is $\pm 2.8\%$. It is derived, following the same methodology as that detailed in ref.~\cite{ATLASLuminosity2013}, from a preliminary calibration of the luminosity scale derived from beam-separation scans performed in November 2012. Events selected by single-electron and single-muon triggers under stable beam conditions and for which all detector subsystems were operational are considered. Un-prescaled single-lepton triggers with different $\pt$ thresholds are combined to increase the overall efficiency. The thresholds are $24$~GeV and $60$~GeV for electron triggers and $24$~GeV and $36$~GeV for muon triggers. The lower threshold triggers include isolation requirements on the candidate leptons, resulting in inefficiencies at higher $\pt$ that are recovered by the triggers with higher $\pt$ thresholds. The trigger isolation criteria are looser than the requirements placed on the final reconstructed leptons. Accepted events are required to have a reconstructed primary vertex with at least five associated tracks with $\pt > 0.4~$GeV. In events with multiple reconstructed vertices the one with the largest sum of the squared $\pt$ of the tracks is taken as the primary interaction vertex.

\section{Monte Carlo simulation}
\label{sec:mcsamples}
Monte Carlo (MC) simulated events are used to help determine SM backgrounds and signal yields in the analysis. Background MC samples are processed through a detector simulation~\cite{Aad:2010ah} based on \geant~\cite{geant4} or a fast simulation using a parameterised response of the showers in the electromagnetic and hadronic calorimeters~\cite{Aad:2010ah}. Additional scale factors are applied to bring the simulation into better agreement with the $2012$ dataset. These include factors for lepton trigger, reconstruction and identification efficiencies. 

Samples of $W$ and $Z/\gamma^*$~\footnote{Hereinafter, all mention of $Z+$jets refers to the $Z/\gamma^{*}$+jets process.} Monte Carlo events with accompanying jets are produced with \sherpa~1.4.1~\cite{Gleisberg:2003xi}, using the {\sc CT10}~\cite{C10} set of parton distribution functions (PDFs). Events generated with \alpgen~2.14~\cite{Mangano:2002ea} use the {\sc CTEQ6L1}~\cite{Pumplin:2002vw} PDF set and are interfaced to {\sc Pythia}~$6.426$~\cite{Pythia} for parton showers and hadronisation with the {\sc Perugia2011C} tune; these \alpgen samples are used to assess modelling uncertainties. The cross-section normalisations are set to the inclusive next-to-next-to-leading-order (NNLO) prediction from the {\sc DYNNLO} program~\cite{DYNNLO}. 

The production of top quark pairs ($t\bar{t}$) is modelled using {\sc POWHEG}~r2129~\cite{powhegPaper} for the matrix element using the {\sc CT10} PDF set, with the top quark mass set to $172.5\GeV$. Parton showering and hadronisation are performed with {\sc Pythia}~6.426 with the {\sc Perugia2011C} tune. Modelling uncertainties are assessed using events generated with \alpgen~2.14, using the {\sc CTEQ6L1}~\cite{Pumplin:2002vw} PDF set and interfaced to \herwig~6.5.20~\cite{herwig} for parton showers and hadronisation. The $t\bar{t}$ cross-section is normalised to $253^{+ 13}_{- 15}$~pb, calculated at NNLO in QCD including resummation of next-to-next-to-leading logarithmic (NNLL) soft gluon terms with {\sc Top++}~2.0~\cite{Cacciari:2011hy, Baernreuther:2012ws, Czakon:2012zr, Czakon:2012pz, Czakon:2013goa, Czakon:2011xx}.

Samples are generated separately for each of the three single-top production modes: $s$-channel, $t$-channel and $Wt$-channel. For the $s$- and $Wt$-channel, events are generated with \mcatnlo~4.06~\cite{mcatnlo}, interfaced to \herwigpp~2.6.3~\cite{Bahr:2008pv} for parton showering and hadronisation. The $t$-channel events are generated with {\sc AcerMC}~3.8~\cite{acermc} interfaced to {\sc Pythia}~6.426. For all three channels, the {\sc CT10} PDF set is used with the {\sc AUET2B}~\cite{AUET2B} tune, and events are reweighted using the NNLO+NNLL cross-sections as given in ref.~\cite{Kidonakis:2010tc,Kidonakis:2010ux,Kidonakis:2011wy}. Diboson ($WW$, $WZ$, $ZZ$) production is simulated with \herwig~6.5.20 using the {\sc CTEQ6L1} PDF set and the {\sc AU2} tune~\cite{AU2}, normalised to the NLO prediction of {\sc MCFM} $6.2$~\cite{Campbell:1999ah,Campbell:2011bn}.

The canonical Monte Carlo generators for the production of black hole signals are \charybdis~2.104~\cite{Frost:2009cf} and \blackmax~2.2.0~\cite{Dai:2007ki,Dai:2009by}. Both programs are able to simulate a range of rotating and non-rotating black hole and string ball states, exploring the theoretical modelling uncertainties discussed in section~\ref{sec:introduction}. A variety of potential black hole signals simulated with both generators are used to illustrate possible black hole models. They are described in detail below and summarised in table~\ref{tab:signalsConsidered}. The shower evolution and hadronisation of all signal samples uses {\sc Pythia}~8.165~\cite{Pythia8}, with the {\sc MSTW2008LO~\cite{Martin:2009iq}} PDF set and the {\sc AU2} tune. The mass of the black hole is used as the factorisation and renormalisation scales. The detector response is simulated using the ATLAS fast simulation~\cite{ATLAS:1300517}. The benchmark event samples are generated for two, four and six extra dimensions.

Both Monte Carlo generators are able to include the effects of the black hole angular momentum, with similar treatments of the Hawking evaporation. Moreover, they contain complementary and different modelling options for the more uncertain decay phases. 
Both generators model losses of mass and angular momentum in the production phase: \charybdis uses a model based on the Yoshino--Rychkov bounds~\cite{Yoshino:2005hi,Frost:2009cf}, favouring smaller losses of mass and angular momentum in the form of gravitons, whereas \blackmax parameterises the losses as fixed fractions of their initial-state values. 
For each generator, a benchmark model including these loss models is used to investigate their effect. The \blackmax sample assumes a 10\% loss into photon modes.
\blackmax can also model graviton emission in the non-rotating case, which is considered in another benchmark sample. The modelling of the remnant phase can have large effects on the event multiplicity, and hence the experimental signature. \blackmax uses a final-burst remnant model, which gives high-multiplicity remnant states~\cite{Dai:2007ki}; \charybdis benchmarks are generated with both low- and high-multiplicity remnant decays, corresponding to fixed two-body decay, and variable decay with a mean of four particles, respectively.
The high-multiplicity options of both generators produce similar distributions of particle multiplicities and $\pt$. Baryon and lepton numbers may not be conserved in black hole interactions~\cite{Hawking:2005kf,Hawking:2014tga}; however, both generators conserve baryon number to avoid problems with colour in hadronisation. The default generator treatment is to violate lepton number, though both options are available. A benchmark sample with lepton number conservation is produced with \blackmax, for $n=4$ only. String ball samples are produced with \charybdis for both rotating and non-rotating cases, six extra dimensions, a string coupling $g_\mathrm{S} = 0.4$, and $\MD = g_\mathrm{S}^{-2/(n+2)} \MS = 1.26~\MS$. 

For each benchmark model, samples are generated with $\MD$ varying from $1.5$ to $4$~TeV ($\MS$ varying from $1$ to $3$~TeV for string ball models) and $\MTH$ from $4$--$6.5$~TeV, so as to cover the production cross-sections to which the current data are sensitive. The productions cross-sections are calculated by the event generators.

\begin{table}[htb]
	\begin{center}
		\begin{tabular}{|c|c|c|c|}\hline
			Generator & Angular Mom. & Description & $n$ considered\\ \hline

			\charybdis & Non-rotating & Black holes: High multiplicity remnant  &  2, 4, 6 \\ 
			& Rotating & Black holes: High multiplicity remnant  &  2, 4, 6 \\
			& Rotating & Black holes: Low multiplicity remnant  &  2, 4, 6 \\
			& Rotating & Production loss model (gravitons)  &  2, 4, 6 \\ \hline
			\charybdis & Non-rotating & String balls   &   6 \\
			& Rotating & String balls & 6 \\ \hline
			{\sc BlackMax}  & Non-rotating & Black holes: High multiplicity remnant  &  2, 4, 6 \\ 
			& Rotating & Black holes: High multiplicity remnant  &  2, 4, 6\\
			& Non-rotating & Black holes with graviton  &  2, 4, 6  \\
			& Rotating & 10\% Production loss model (photons) &  2, 4, 6 \\ 
			& Rotating & Lepton number conservation  &  4 \\ \hline
			\hline
		\end{tabular}
		\caption{Summary of the TeV-scale gravity benchmark models considered. }
		\label{tab:signalsConsidered}
	\end{center}
\end{table}

%%%%%%%%%%%%%%%%%%%%%%%%%%%%%%%%%%%%%%%%%%%%%%%%%%%%%%%%%%%%%%%%%%%%%%%%%%%%%%
\section{Object reconstruction}
\label{sec:objectreconstruction}

Jets are reconstructed using the anti-$k_t$ jet clustering algorithm~\cite{Cacciari:2008gp} with radius parameter $R = 0.4$. The inputs to the jet algorithm are clusters seeded from calorimeter cells with energy deposits significantly above the measured noise~\cite{Lampl:2008}.
Jet energies are corrected~\cite{Aad:2011he} for detector inhomogeneities, and the non-compensating response of the calorimeter, using factors derived from test beam, cosmic ray and $pp$ collision data, and from the full detector simulation. Furthermore, jets are corrected for energy from additional $pp$ collisions (pile-up) using a method proposed in ref.~\cite{Aad:2014fxa}, which estimates the pile-up activity in any given event, as well as the sensitivity of any given jet to pile-up. Selected jets are required to have $\pt > 60 \GeV$ and $|\eta| < 2.8$. Events containing jets failing to satisfy the quality criteria that discriminate against electronic noise and non-collision backgrounds are rejected~\cite{Aad:2011he}.

Electrons are reconstructed from clusters in the electromagnetic calorimeter associated with a track in the ID~\cite{Aad:2011mk}, with the criteria re-optimised for 2012 data. Two sets of electron identification criteria based on the calorimeter shower shape, track quality and track
matching with the calorimeter cluster are referred to as ``medium'' and ``tight'', with ``tight'' offering increased background rejection over ``medium" in exchange for some loss in identification efficiency. Electrons are required to have $\pt > 60 \GeV$, $|\eta|< 2.47$ and to satisfy the ``medium'' electron definition. Candidates in the transition region between barrel and end-cap calorimeters, $1.37< |\eta| < 1.52$, are excluded.
Electron candidates are required to be isolated:  the sum of the $\pt$ of tracks within a cone of size $\Delta R = \sqrt{(\Delta \eta)^2 + (\Delta \phi)^2} = 0.2$ around the electron candidate is required to be less than $10\%$ of the electron $\pt$.

Muon tracks are reconstructed from track segments in the various layers of the muon spectrometer and then matched to corresponding tracks in the inner detector~\cite{Aad:2014zya}. In order to ensure good $\pt$ resolution, muons are required to have at least three hits in each of the layers of either the barrel or end-cap region of the MS, and at least one hit in two layers of the trigger chambers. Muon candidates passing through known misaligned chambers are rejected, and the difference between the independent momentum measurements obtained from the ID and MS must not exceed five times the sum in quadrature of the uncertainties of the two measurements. Each muon candidate is required to have a minimum number of hits in each of the subsystems of the ID, and to have $\pt>60$~GeV and $|\eta| < 2.4$. In order to reject muons resulting from cosmic rays, requirements are placed on the distance of each muon track from the primary vertex:
$|z_0| < 1$~mm and $|d_0| < 0.2$~mm, where $z_0$ and $d_0$ are the impact parameters of each muon in the longitudinal direction and transverse plane, respectively.  To reduce the background from non-prompt sources such as heavy-flavour decays, muons must be isolated: the $\pt$ sum of tracks within a cone of size $\Delta R = 0.3$ around the muon candidate is required to be less than $5\%$ of the muon $\pt$. Ambiguities between the reconstructed jets and leptons are resolved by applying the following criteria: jets within a distance of $\Delta R = 0.2$ of an electron candidate are rejected; furthermore, any lepton candidate with a distance $\Delta R < 0.4$ to the closest remaining jet is discarded.

The signal selection places no requirement on whether or not selected jets originate from the hadronisation of a $b$-quark ($b$-jets). However, $b$-jets are used in the definition of control regions, either by requiring at least one $b$-tagged jet, or by vetoing any event with at least one $b$-tagged jet.
To identify $b$-jets, the employed algorithm~\cite{btagging} uses multivariate techniques to combine information derived from tracks within jets, such as impact parameters and reconstructed vertices displaced from the primary vertex. The efficiency of tagging a $b$-jet in simulated $t\bar{t}$ events is estimated to be $70\%$, with charm jet, light-quark jet and $\tau$ lepton rejection factors of about $5$, $147$ and 13, respectively. Scale factors associated with the identification efficiencies of $b$-jets are applied to bring the simulation into better agreement with the data~\cite{ATLAS-CONF-2014-004}.

The missing transverse momentum $\vec{p}_{\mathrm{T}}^{\,\mathrm {miss}}$, with its magnitude $\met$, is defined as the negative vectorial $\pt$ sum of reconstructed objects in the event, comprising selected leptons, jets with $\pt > 20 \GeV$, any additional non-isolated muons with $\pt > 10~\GeV$, and calorimeter clusters not belonging to any of the aforementioned object types~\cite{Aad:2012re}. $\met$ is only used to define control regions for the background estimation and not to define the signal region. The transverse mass, $m_\mathrm{T}$, is also used in the definition of control regions. It is calculated from the lepton transverse momentum vector, $\vec{p}_\mathrm{T}^{\,\ell}$, and the missing transverse momentum vector $\vec{p}_{\mathrm{T}}^{\,\mathrm {miss}}$:
\begin{equation}
 m_\mathrm{T} = \sqrt{2 \cdot \pt^{\,\ell} \cdot \met \cdot ( 1 - \cos(\Delta \phi ( \vec{p}_\mathrm{T}^{\,\ell} , \vec{p}_{\mathrm{T}}^{\,\mathrm {miss}} ) ) )} \> \> .
\end{equation}

%%%%%%%%%%%%%%%%%%%%%%%%%%%%%%%%%%%%%%%%%%%%%%%%%%%%%%%%%%%%%%%%%%%%%%%%%%%%%%
\section{Event selection}
\label{sec:eventselection}

The selected events contain at least one high-$\pt$ isolated lepton and at least two additional objects (leptons and jets). Two statistically independent channels are defined, based on whether the highest-$\pt$ lepton matching a lepton reconstructed by the trigger is an electron or a muon. This lepton is called the ``leading" lepton. For the electron channel, the leading electron is required to pass the ``tight" selection criteria. The muon channel has a lower acceptance, due to the stringent hit requirements in the muon spectrometer.

\begin{figure*}[htb]
\begin{center}
\includegraphics[width=0.45\textwidth]{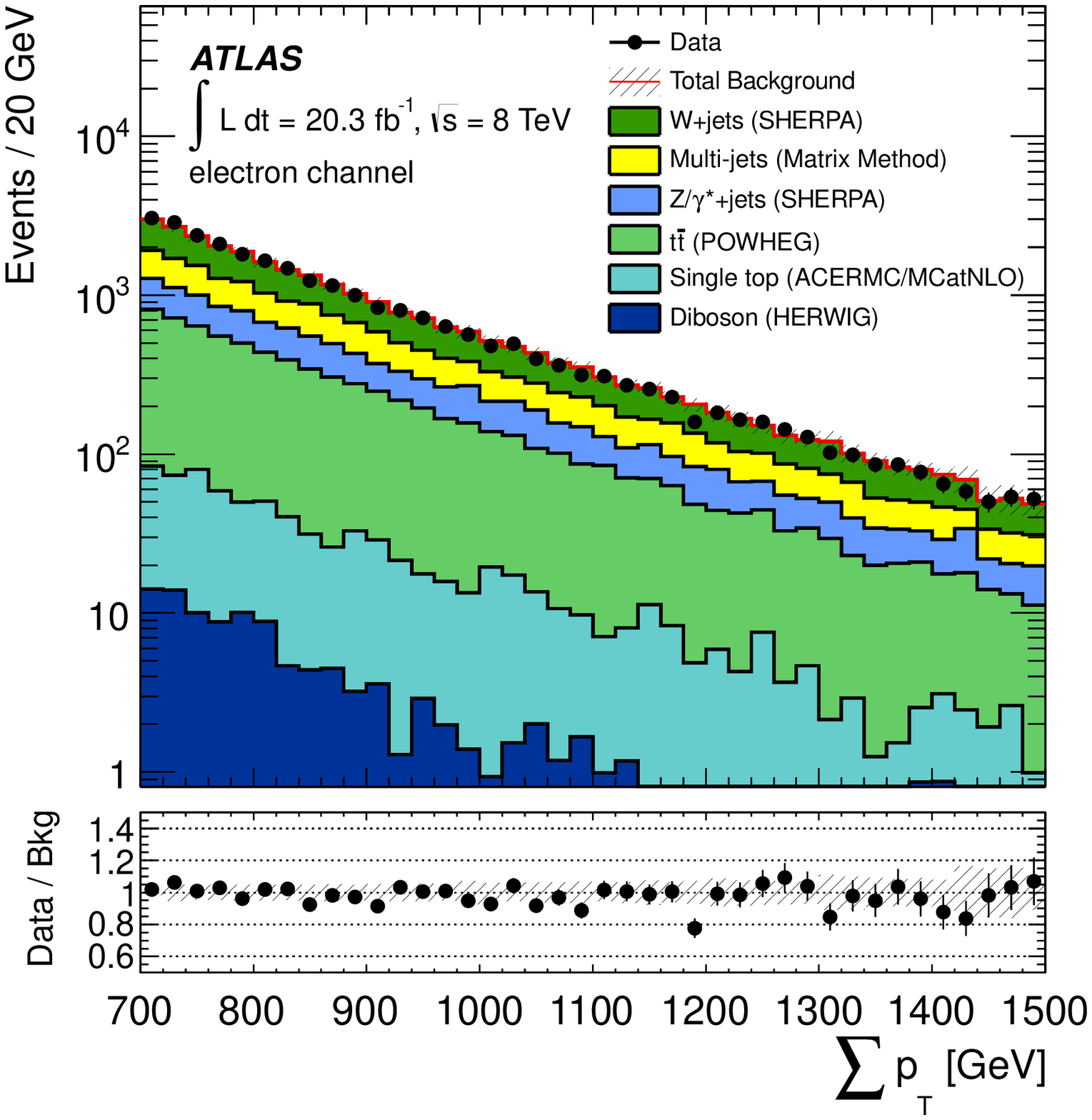}
\includegraphics[width=0.45\textwidth]{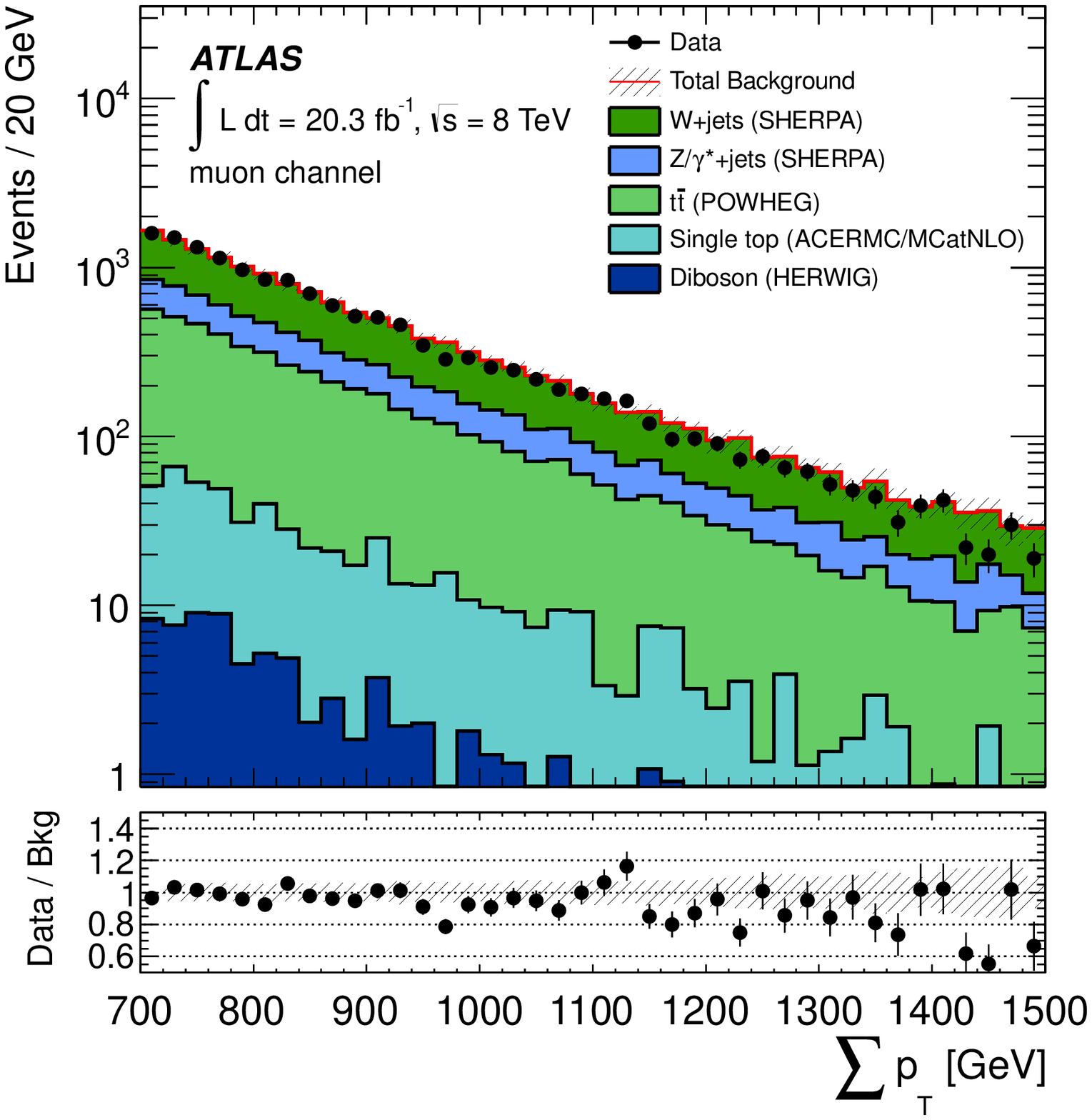}
\caption{The $\sumpt$, after event preselection, in the electron (left) and muon (right) channels. The Monte Carlo distributions are rescaled using scale factors derived in the appropriate control regions, as described in the text. \ratioDescription}
\label{fig:prelep} 
\end{center}
\end{figure*}

The high multiplicity final states of interest are distinguished from SM background events using the quantity:

\begin{equation}
\sumpt \, = \sum_{i=\mathrm{objects}} p_{\mathrm{T},i}\; \textrm{~if }~p_{\mathrm{T},i} > 60 \textrm{~GeV},
\end{equation}
the scalar sum of the transverse momenta of the selected leptons and jets with $\pt > 60$~GeV, described in section~\ref{sec:objectreconstruction}. Events with $700$~GeV $< \sumpt < 1500$~GeV constitute a preselection sample from which special control regions are defined by adding other selection criteria.  Figure~\ref{fig:prelep} shows the $\sumpt$ distribution for preselected events, for the electron and muon channels. The signal, containing multiple high-$\pt$ leptons and jets, would manifest itself as an excess of events at higher $\sumpt$, and is entirely negligible in the preselection region.

For the signal region, in order to reduce the SM background contributions, events are required to contain at least three reconstructed objects with $\pt >100$~GeV, at least one of which must be a lepton, as well as to have a $\sumpt$ of at least $2000$~GeV. In each of the channels, the signal region above $\sumpt=2000$~GeV is divided into multiple slices, with $\sumpt$ thresholds increasing in steps of $200\GeV$. This allows the analysis to be sensitive to a wide range of signal models, and values of \MD and \MTH. 
Events in the range $1000$~GeV $< \sumpt < 2000$~GeV, but otherwise with the same requirements as the signal region, constitute a ``sideband'' region. The contributions from signal models not yet excluded by earlier analyses to the sideband region are well below 1\%. The selection criteria for the sideband and signal regions are summarised in table~\ref{tab:SBSRregionDefinitions}.

\section{Background estimation}
\label{sec:backgrounds}
The dominant sources of Standard Model background in this search are the production of $W$ and $Z$ bosons in association with jets, $t\bar{t}$ production and multi-jet processes.
There are three sources of leptons that can contribute to the background.  
Firstly, the leptonic decays of $W$ and $Z$ bosons and top quarks produce events with real leptons, with associated high-$\pt$ jets (hereinafter called ``prompt'' backgrounds). Secondly, leptons can arise from semileptonic decays of heavy flavour hadrons. These are typically non-prompt and not isolated. Thirdly, other objects such as jets can be misidentified.

The backgrounds are estimated using a combination of data-driven and MC-based techniques. Prompt backgrounds are estimated using MC samples, normalised in data control regions that are dominated by a single background component and kinematically close to the signal region. The multi-jet contribution is estimated using a data-driven technique that is more reliable than simulation for determining ``fake" lepton backgrounds, due to its independence from MC modelling uncertainties such as hadronisation and detector simulation.

At very high $\sumpt$, the number of events in the simulated samples becomes small, and therefore subject to large statistical fluctuations. Therefore, for each background component, the $\sumpt$ distribution is fitted to a functional form to smooth the backgrounds and extrapolate them to very high $\sumpt$.

\begin{table}
	\begin{center}
		\begin{tabular}{|c||c|c|}
			\hline
			\multirow{2}{*}{\textbf{Quantity}} & \multicolumn{2}{|c|}{\textbf{Region}} \\ \cline{2-3}
			 & \textbf{Sideband} & \textbf{Signal}  \\ \hline \hline
			
			$\sumpt$ &  $1000$--$2000$~GeV &  $>$~$2000$~GeV  \\ \hline
			Object multiplicity &  \multicolumn{2}{c|}{at least 3 objects above 100 GeV} \\ \hline
			Leading lepton & \multicolumn{2}{c|}{at least 1 lepton with $\pt > 100$ GeV} \\ \hline 
			 
		\end{tabular}
	\end{center}
	\caption{Definitions of the sideband and signal regions.}
	\label{tab:SBSRregionDefinitions}
\end{table}
\begin{table}
	\begin{center}
		\begin{tabular}{|c||c|c|c|}
			\hline
			{\textbf{Quantity}} & \multicolumn{3}{|c|}{\textbf{Control Region}} \\ \cline{2-4}
			 & $Z+$jets & $W+$jets & {$t\bar{t}$} \\ \hline
			
			$\sumpt$ &  \multicolumn{3}{c|}{$700$--$1500$~GeV} \\ \hline
			Object multiplicity & \multicolumn{3}{c|}{at least 3 objects (leptons or jets) with $\pt > 60$ GeV } \\ \hline
			Leading lepton & \multicolumn{3}{c|}{at least 1 lepton with $\pt > 60$ GeV } \\ \hline \hline
			$m_{\ell \ell}$ & 80--100 GeV & \multicolumn{2}{c|}{n/a}  \\ \hline
			$\met$ & n/a & $> 60$~\GeV & \multicolumn{1}{c|}{n/a } \\ \hline
			\multirow{2}{*}{Lepton multiplicity} & exactly 2, opposite sign & \multicolumn{2}{c|}{\multirow{2}{*}{exactly 1}}  \\
			 & same flavour & \multicolumn{1}{c}{} & \multicolumn{1}{c|}{} \\ \hline
			
			$b$-jet multiplicity &  n/a & exactly 0 & \multicolumn{1}{c|}{$> 1$} \\ \hline
			Jet multiplicity &  \multicolumn{2}{c|}{n/a} & \multicolumn{1}{c|}{$> 3$} \\ \hline
			 
		\end{tabular}
	\end{center}
	\caption{Definitions of the SM background-dominated control regions. % as well as the real and fake enhanced regions used in the data-driven multi-jet estimate. 
The first three rows represent the preselection criteria.}
	\label{tab:regionDefinitions}
\end{table}

\subsection{Prompt background estimation from control regions}
\label{subSec:mcStage}
The background estimates for processes involving prompt leptons are based on MC simulations normalised in control regions, each dominated by a single process, as discussed above. The normalisation factors are determined, separately for the electron and muon channels, for the three main backgrounds: $Z+$jets, $W+$jets and $t\bar{t}$. The control regions are defined in table~\ref{tab:regionDefinitions}. For the $Z+$jets control region, events passing preselection requirements are required to contain two electrons or muons of opposite charge and to have dilepton invariant mass between $80$~GeV and $100\GeV$. The $W+$jets control region consists of events with exactly one lepton, no $b$-tagged jets and $\met$ greater than $60\GeV$, where the last two requirements help to reduce the $t\bar{t}$ and $Z+$jets/multi-jet contributions, respectively. The $t\bar{t}$ control region consists of events with exactly one lepton and at least four jets, of which at least two must be tagged as $b$-jets. The final criterion ensures no overlap with the $W+$jet control region and preferentially selects for the top quark decays. The purities of the $Z+$jets, $W+$jets and $t\bar{t}$ control regions are estimated from Monte Carlo simulations to be about $98\%$, $70\%$ and $90\%$, respectively. 

The number of events predicted by the MC simulation is compared to the observed number of events in data in each of the control regions, to derive the scale factors used to normalise the backgrounds. Due to non-negligible contamination by $W+$jets events in the $t\bar{t}$ control region and vice-versa, two coupled equations determine the two normalisations that lead to agreement between data and MC simulation. The derived scale factors to be applied to the background predictions in the electron (muon) channels are 1.00 (1.08) for $t\bar{t}$, 0.76 (0.81) for $W$+jets, and 0.90 (0.93) for $Z$+jets. They are compatible between channels within their statistical uncertainties. 

The much smaller contributions from single-top and diboson processes are estimated to comprise approximately $2\%$ and $0.5\%$, respectively, of the events in the sideband and signal regions. Their estimates are taken directly from Monte Carlo simulations.

\begin{figure}[p]
	\begin{center}
		\subfigure[$Z+$jets CR, dilepton invariant mass, electron channel.]{
		\includegraphics[width=0.38\textwidth]{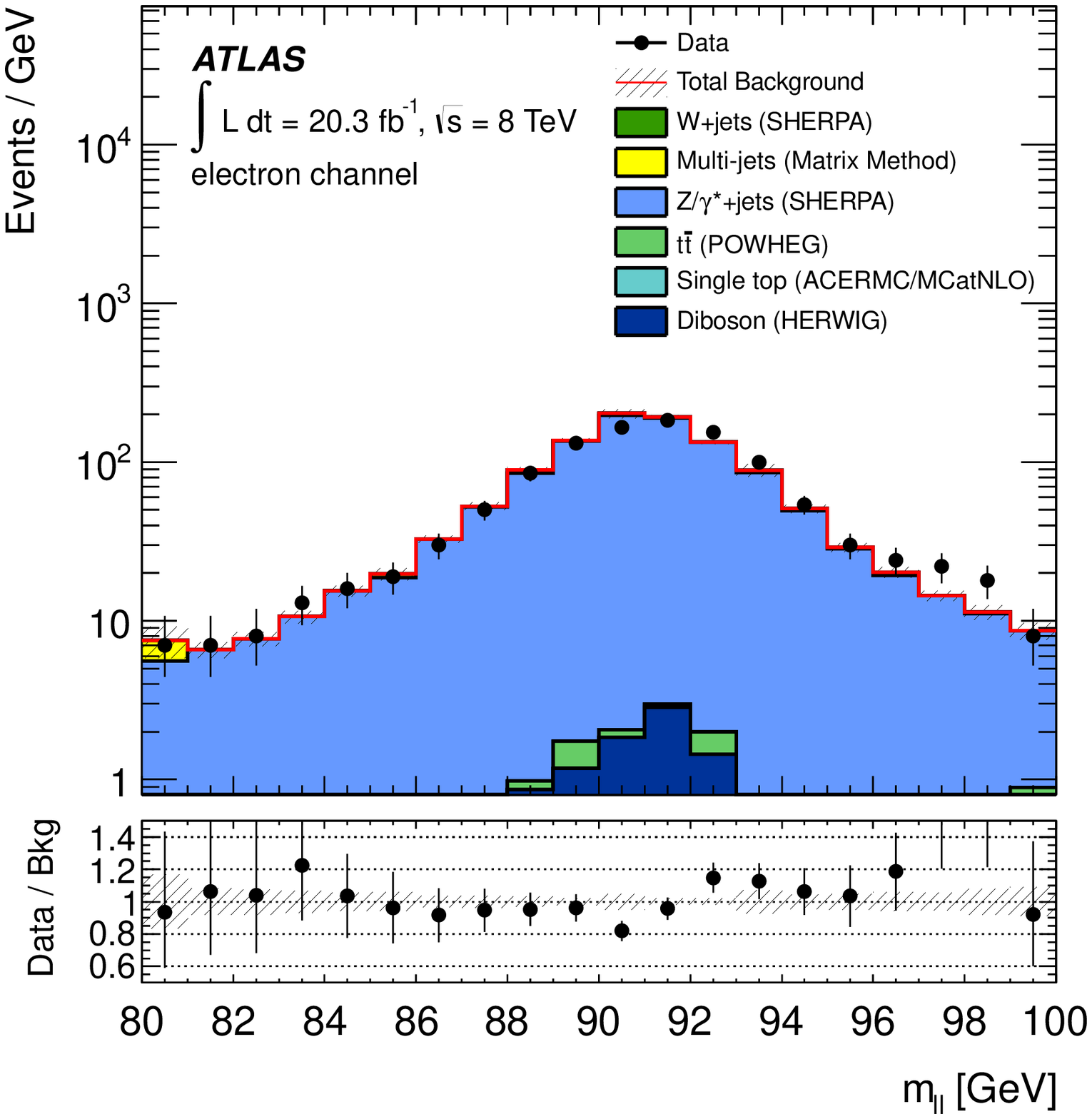}}
		\subfigure[$Z+$jets CR, dilepton invariant mass, muon channel.]{
		\includegraphics[width=0.38\textwidth]{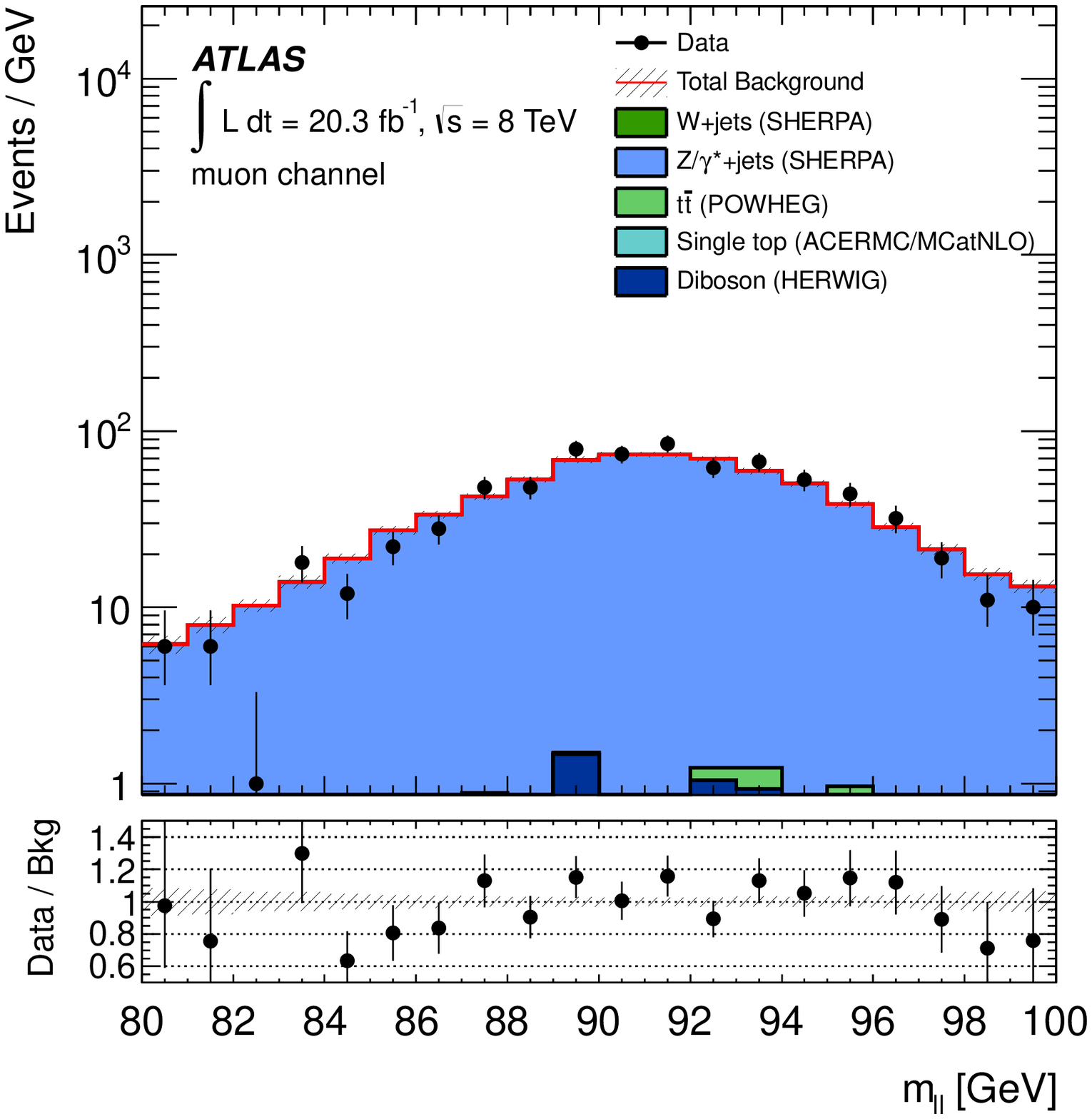}}
		\subfigure[$W+$jets CR, $m_\mathrm{T}$, electron channel.]{
		\includegraphics[width=0.38\textwidth]{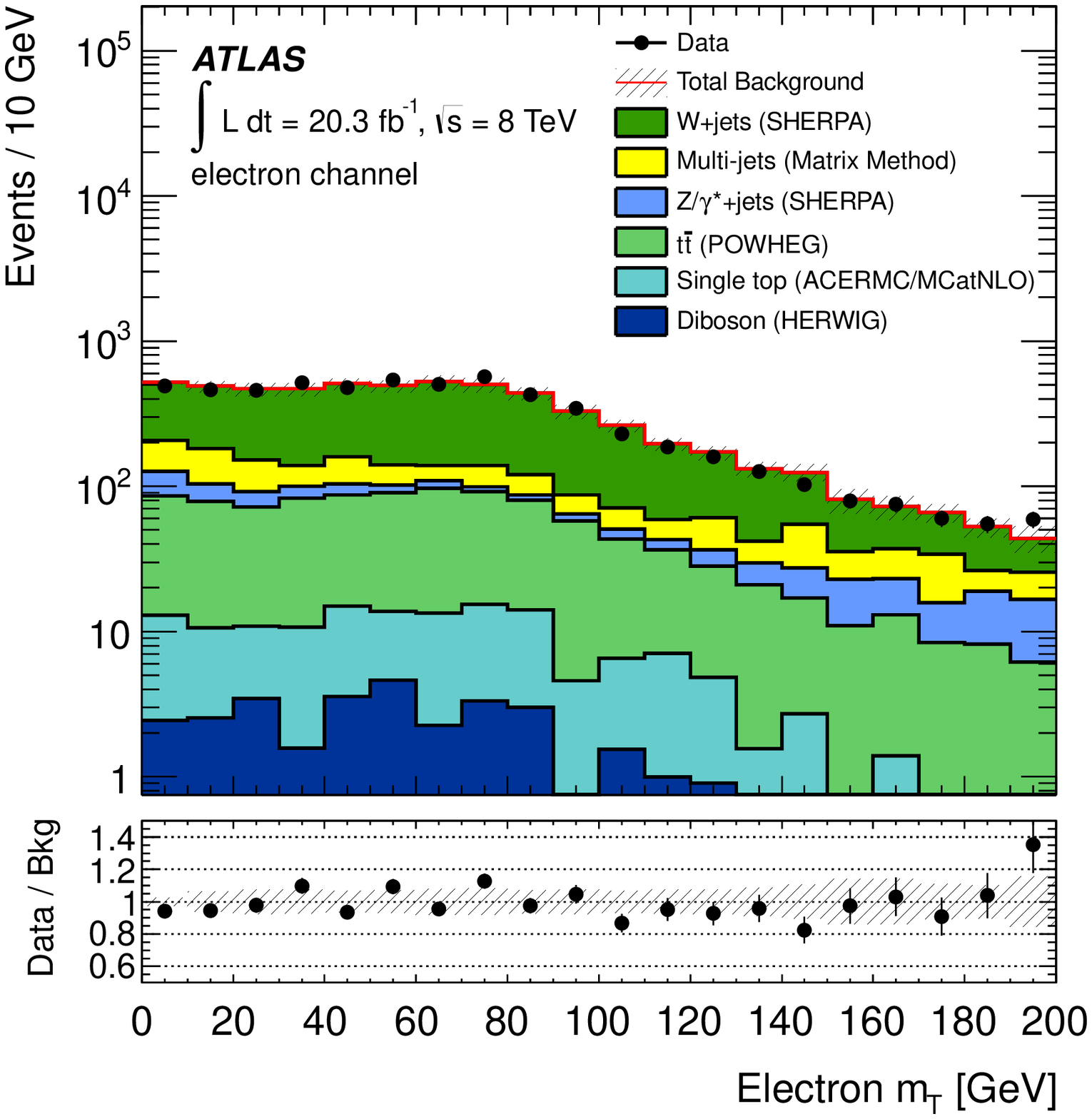}}
		\subfigure[$W+$jets CR, $m_\mathrm{T}$, muon channel.]{
		\includegraphics[width=0.38\textwidth]{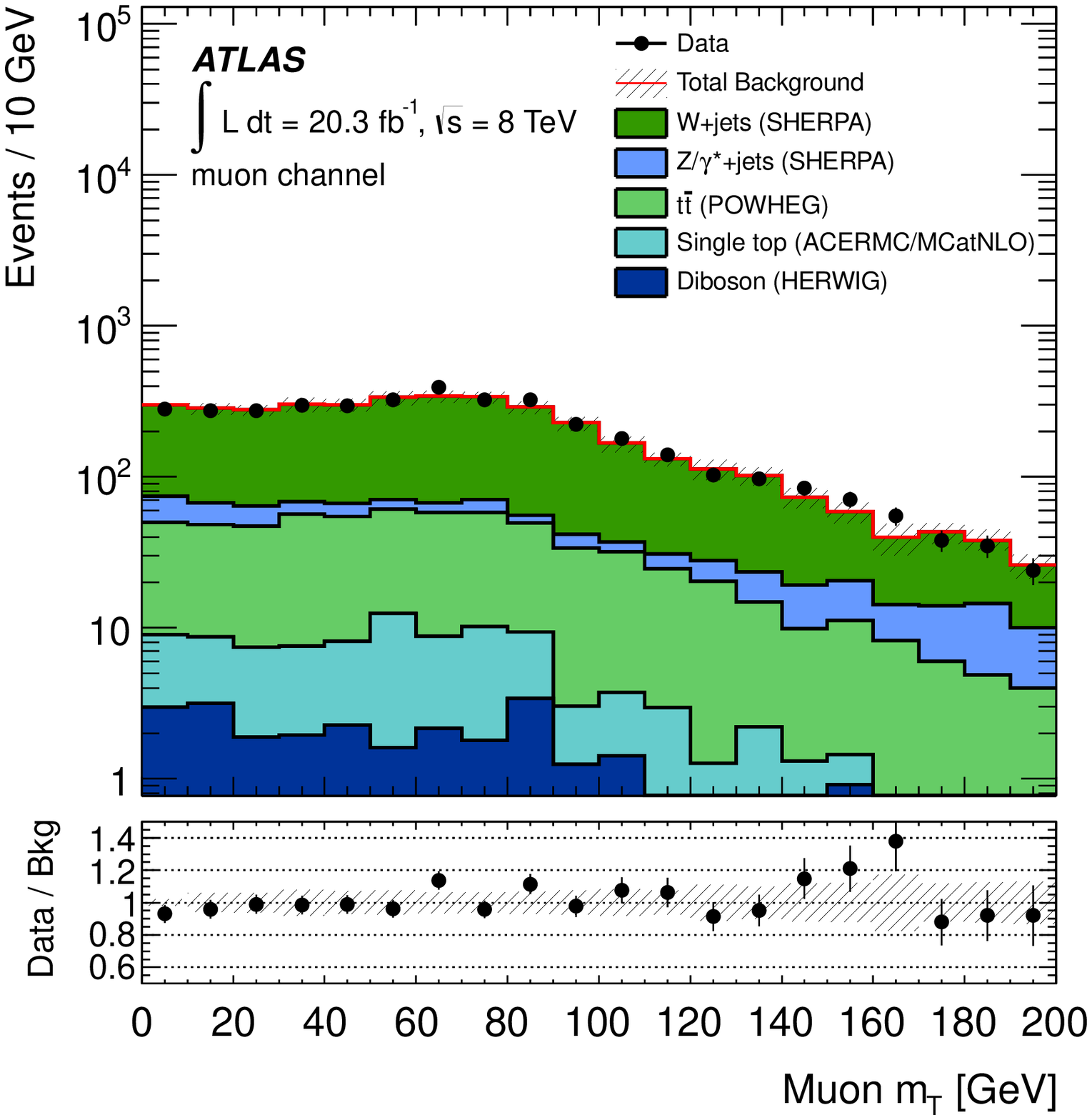}}
		\subfigure[$t\bar{t}$ CR, jet multiplicity, electron channel.]{
		\includegraphics[width=0.38\textwidth]{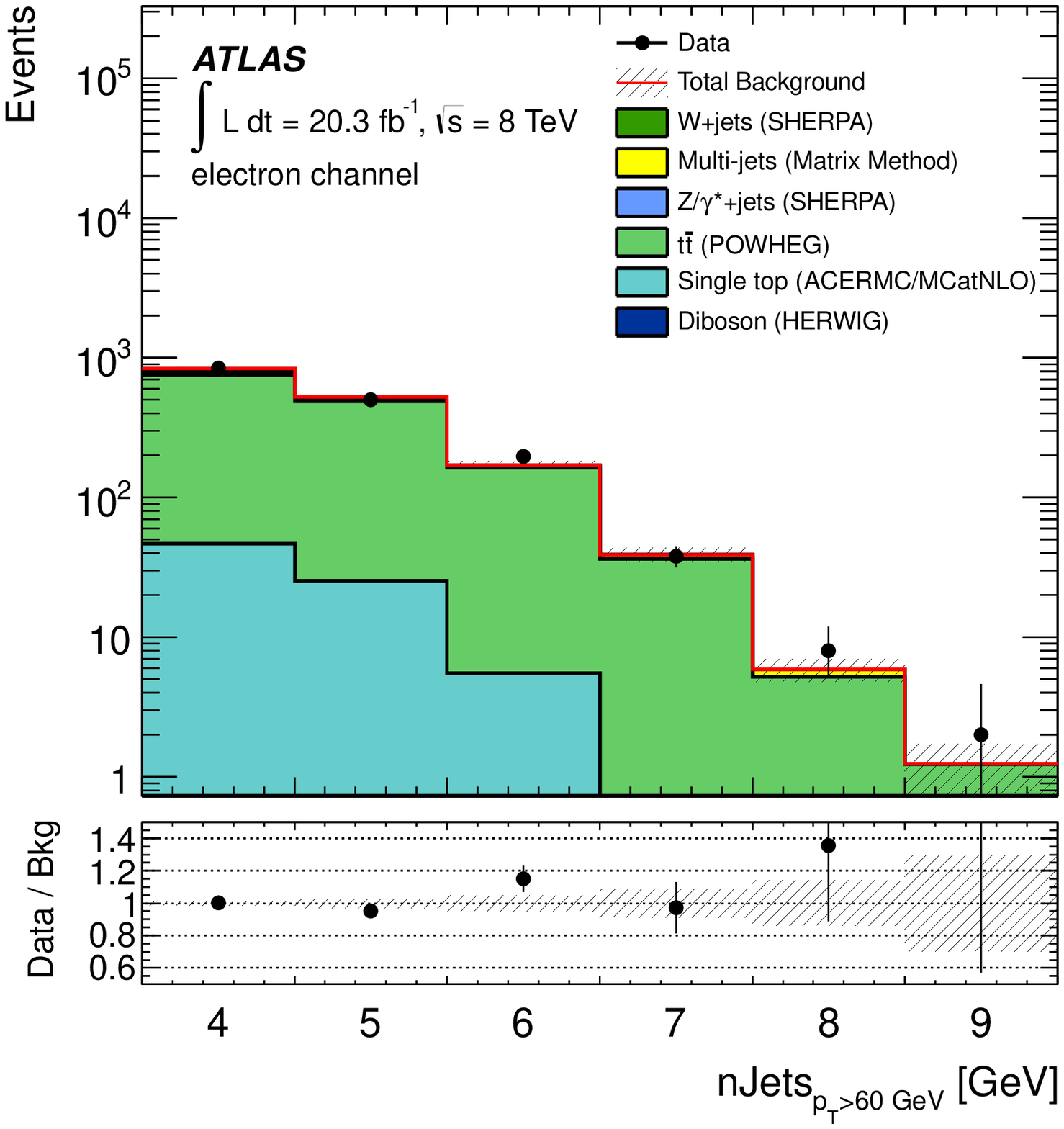}}
		\subfigure[$t\bar{t}$ CR, jet multiplicity,  muon channel.]{
		\includegraphics[width=0.38\textwidth]{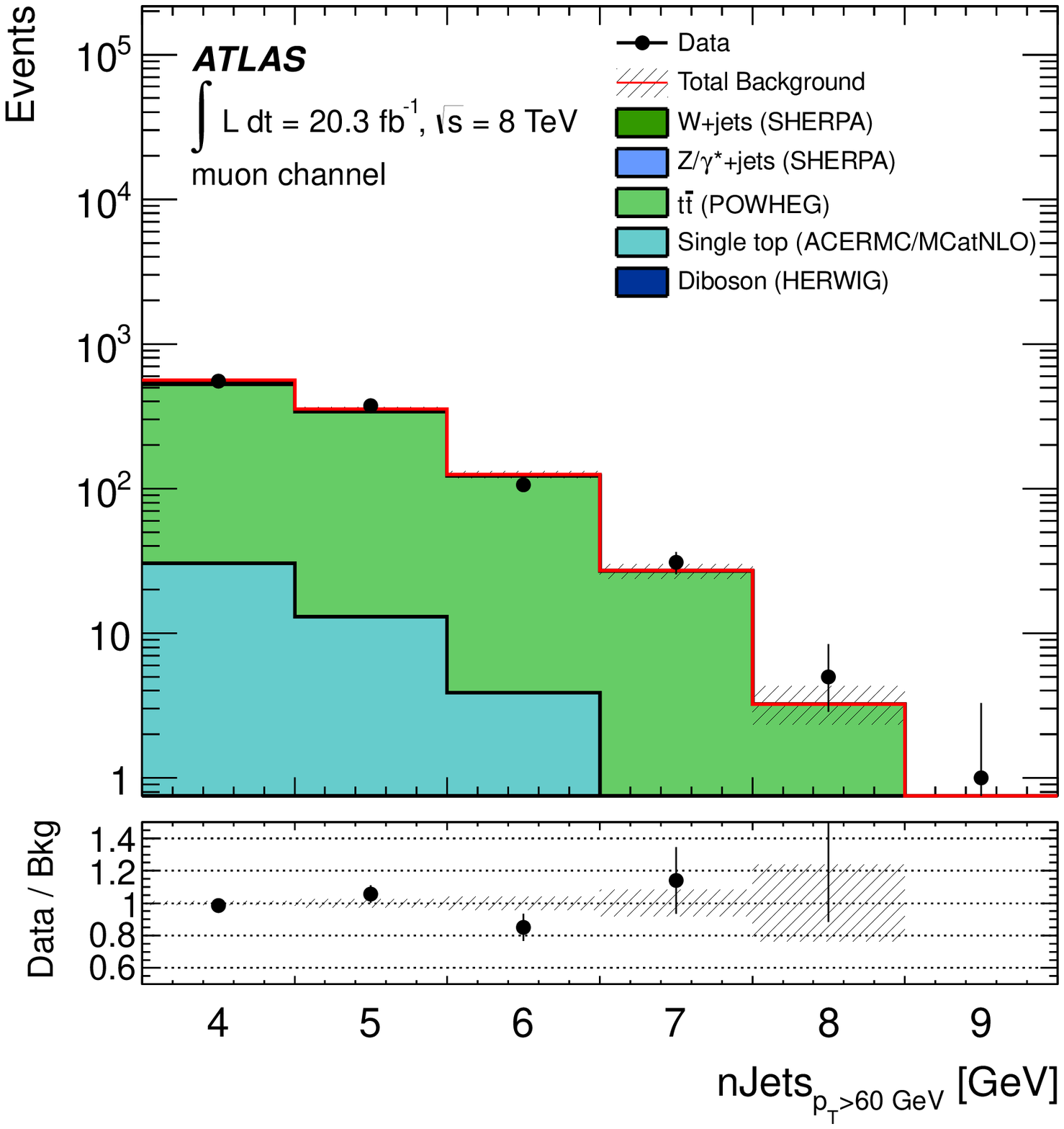}}

		\caption{Kinematic distributions for the three control regions (CR). The Monte Carlo samples are normalised to data using scale factors, according to the method described in section~\ref{sec:backgrounds}. The regions are defined in table~\ref{tab:regionDefinitions}. Some background contributions are very small in specific control regions. \ratioDescription}
		\label{fig:kinematicDistributionsCR}  
	\end{center}
\end{figure}

\begin{figure}[p]
	\begin{center}
		\subfigure[$Z$+jets CR, \sumpt, electron channel.]{
		\includegraphics[width=0.38\textwidth]{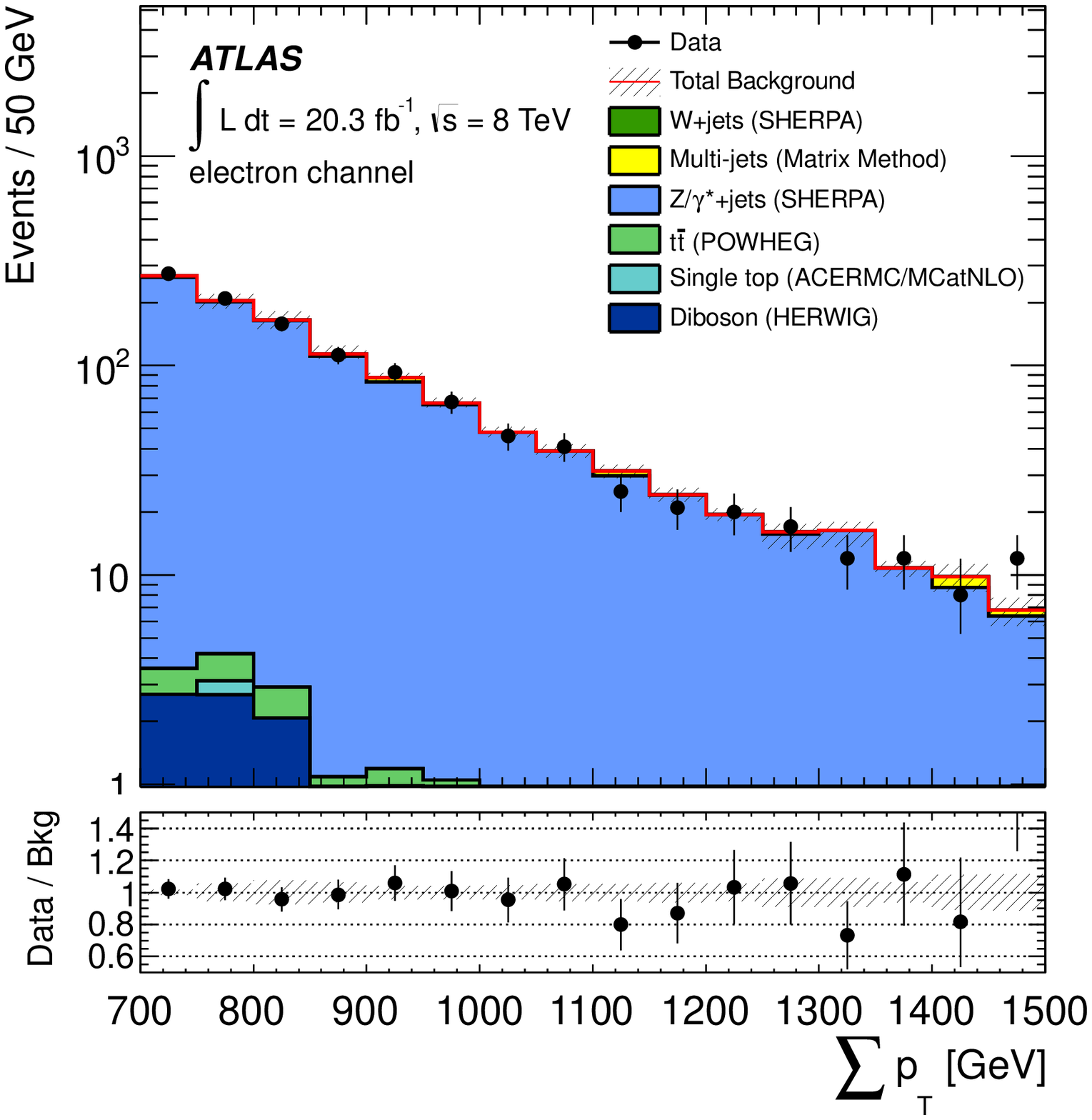}}
		\subfigure[$Z$+jets CR, \sumpt, muon channel.]{
		\includegraphics[width=0.38\textwidth]{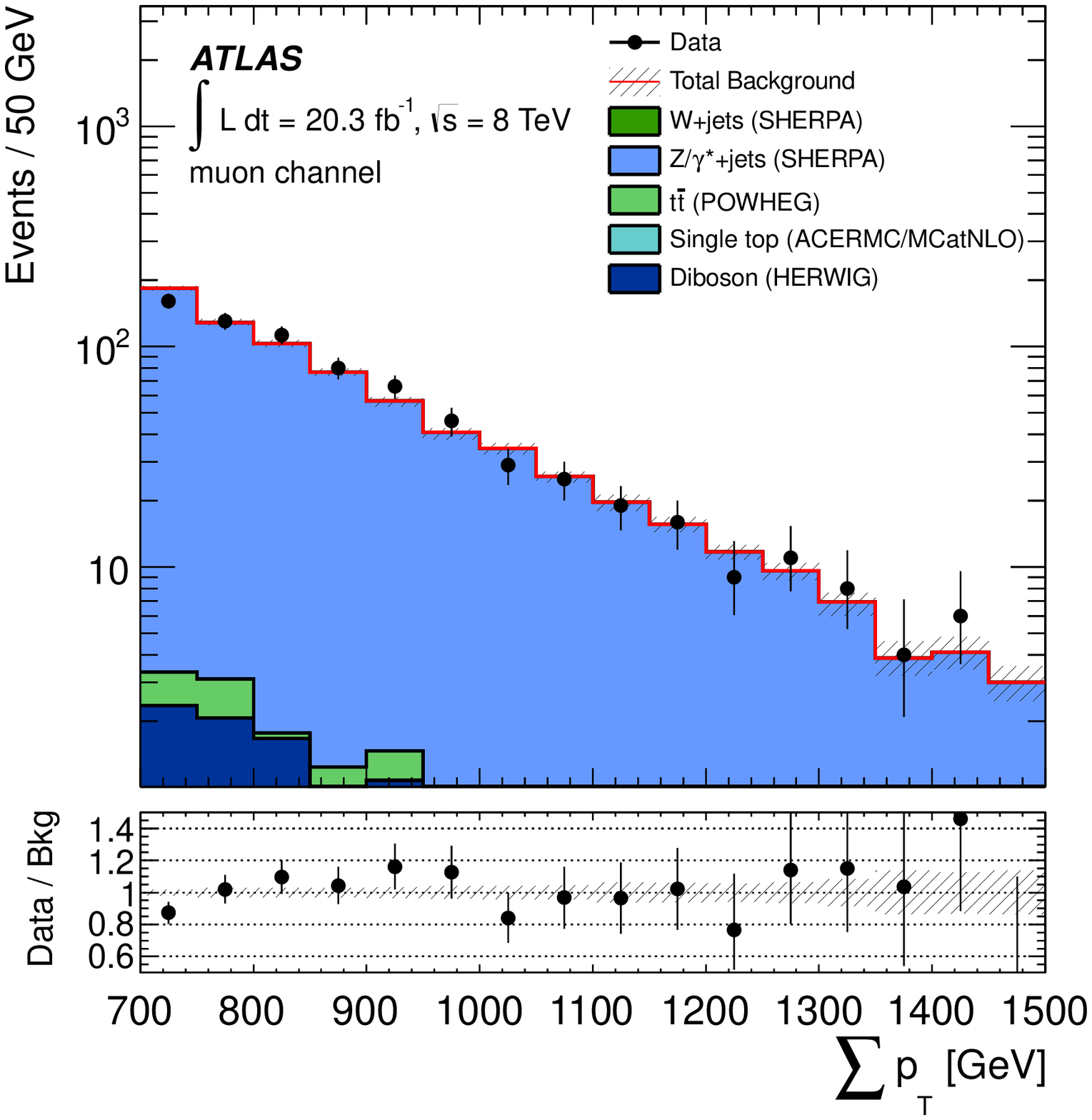}}
		\subfigure[$W$+jets CR, \sumpt, electron channel.]{
		\includegraphics[width=0.38\textwidth]{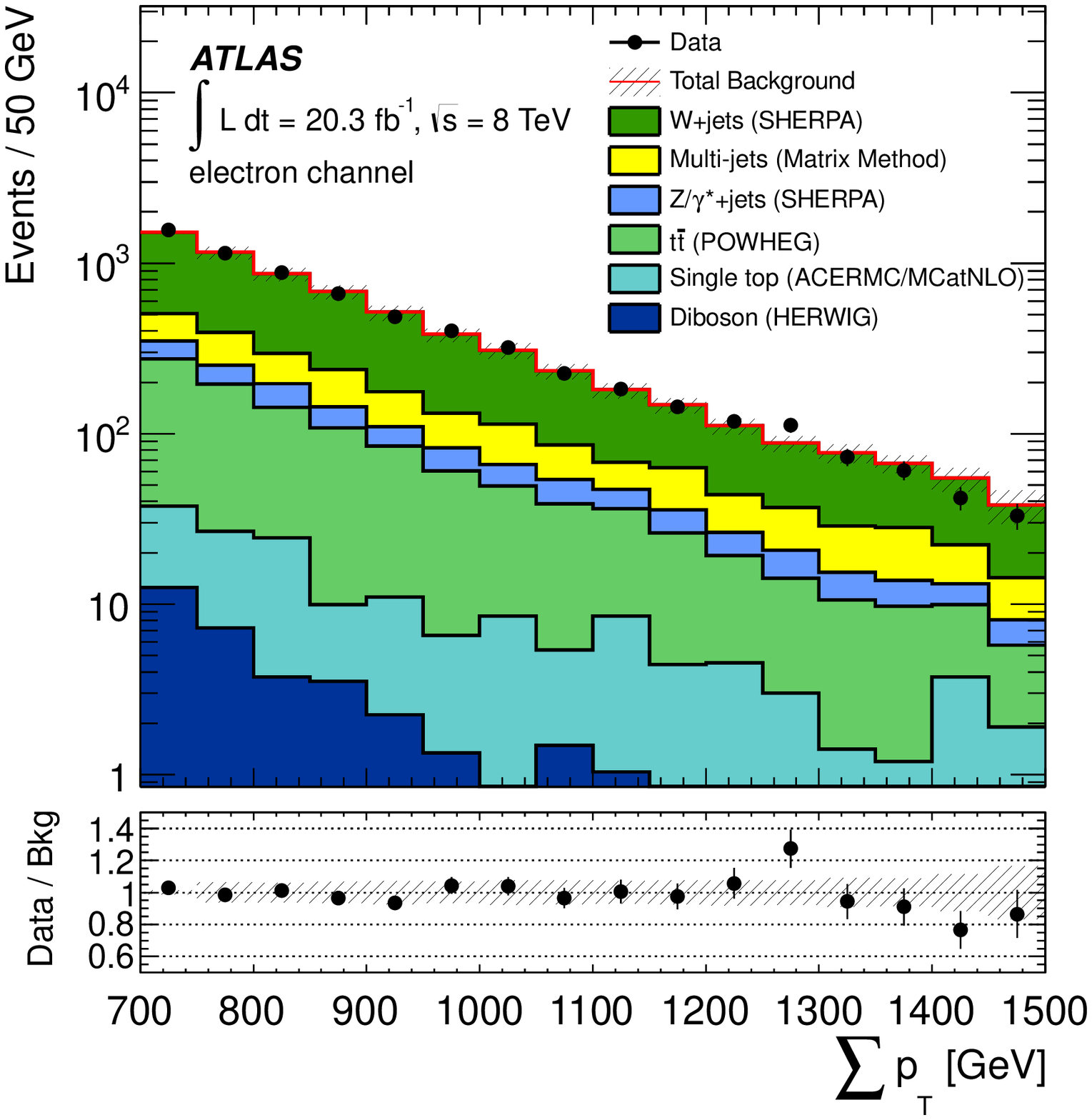}}
		\subfigure[$W$+jets CR, \sumpt, muon channel]{
		\includegraphics[width=0.38\textwidth]{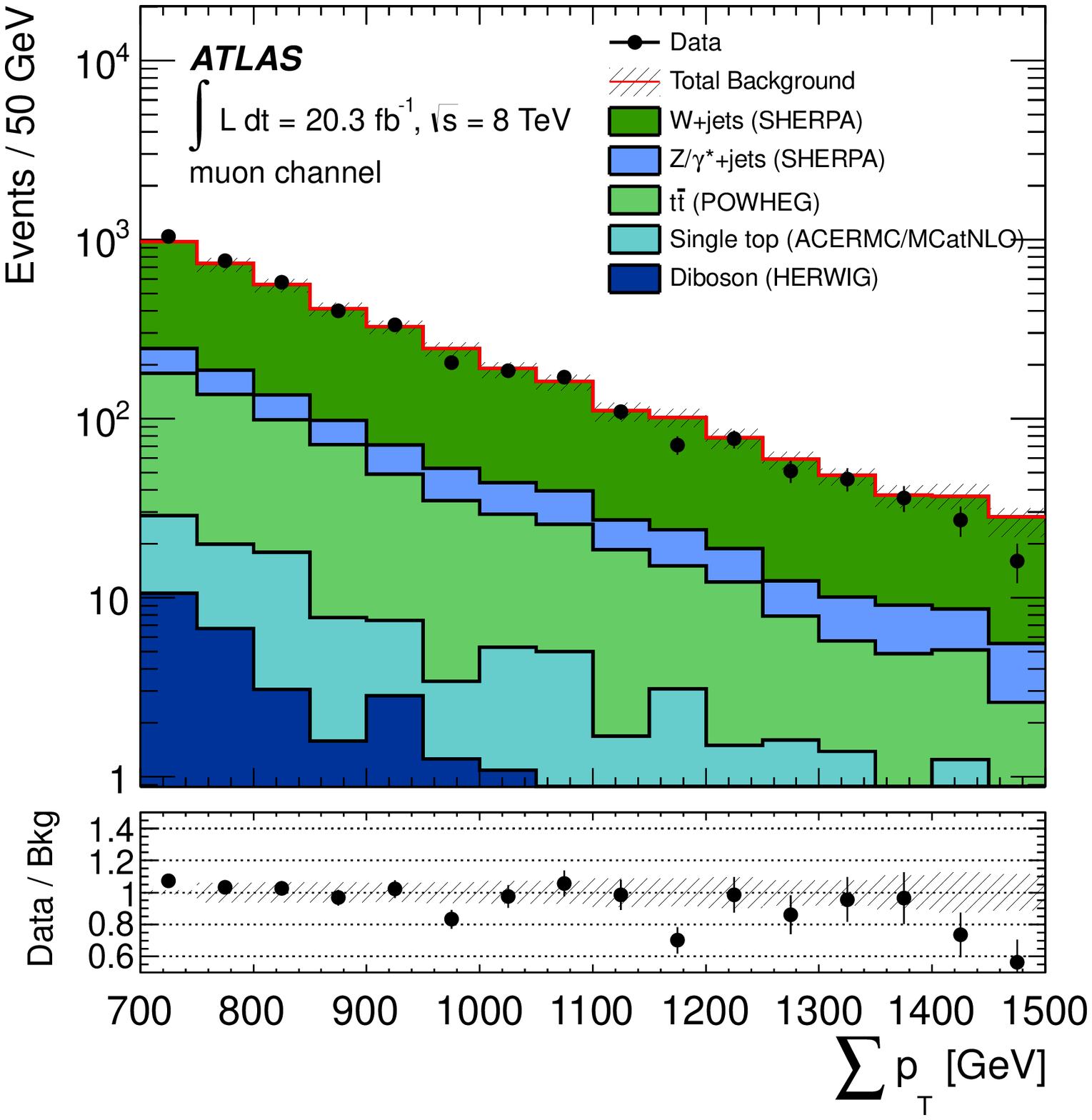}}
		\subfigure[$t\bar{t}$ CR, \sumpt, electron channel]{
		\includegraphics[width=0.38\textwidth]{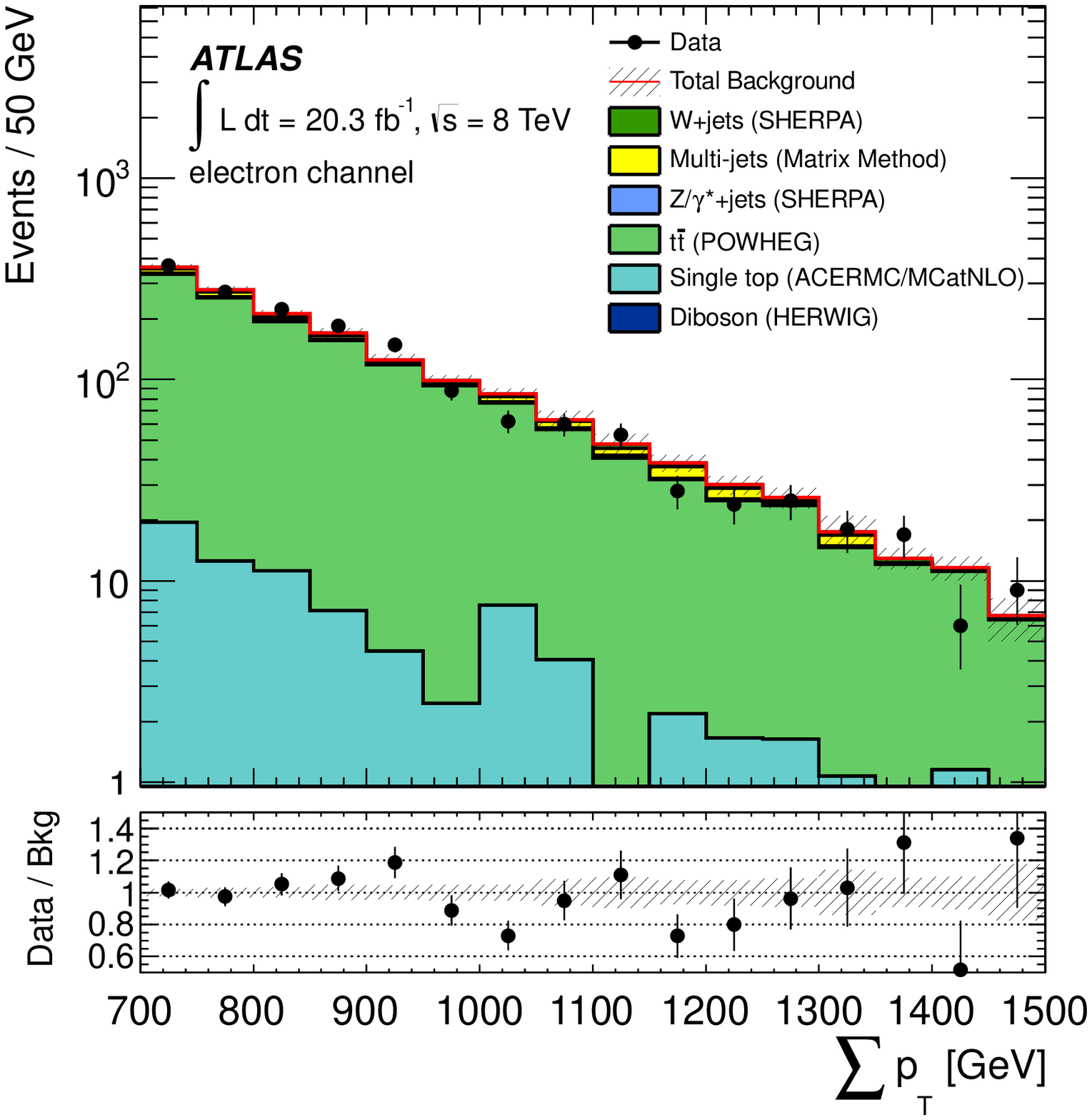}}
		\subfigure[$t\bar{t}$ CR, \sumpt, muon channel.]{
		\includegraphics[width=0.38\textwidth]{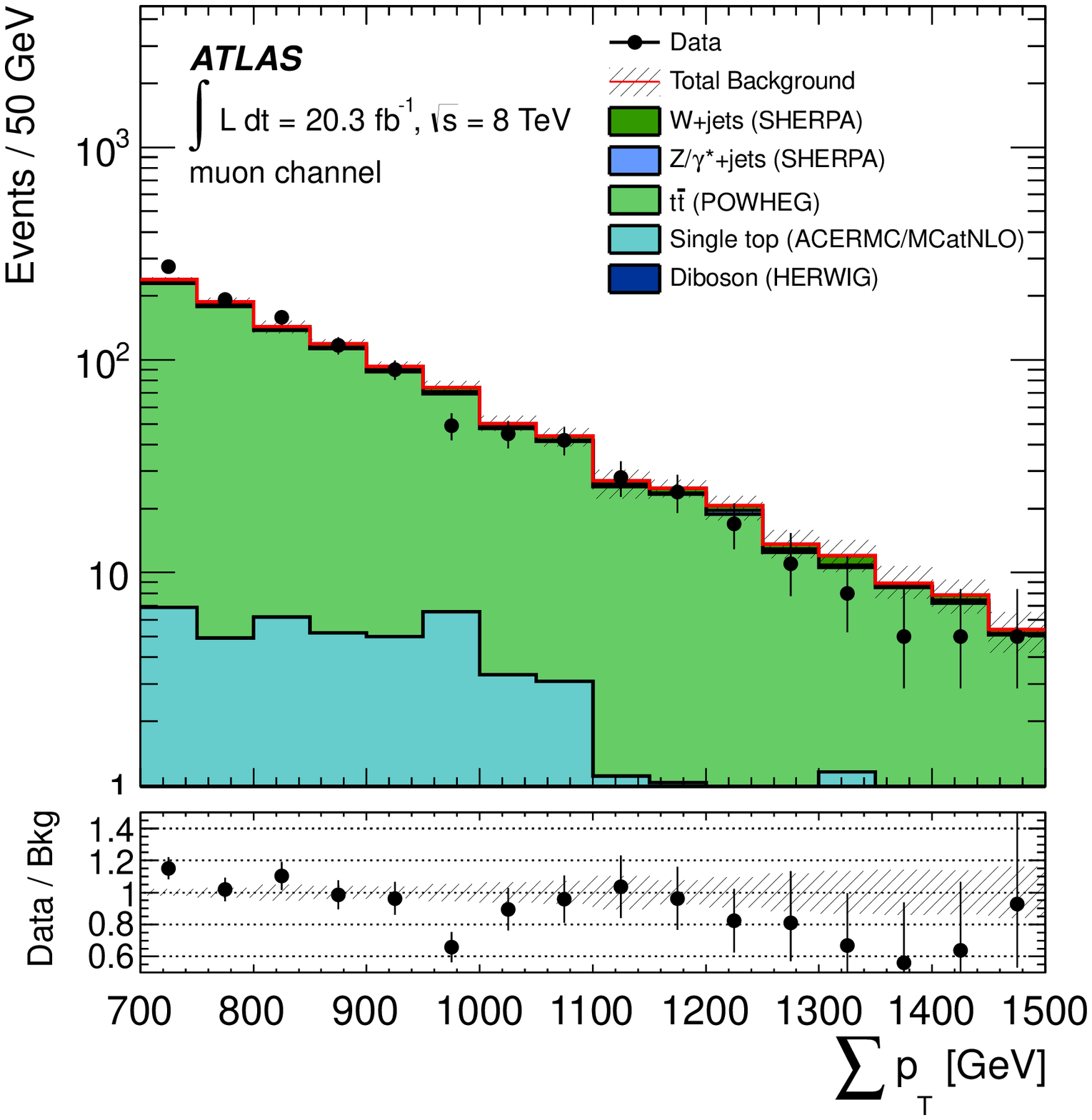}}
		\caption{\sumpt distributions for each control region (CR). The Monte Carlo samples are normalised to data using scale factors, according to the method described in section~\ref{sec:backgrounds}. The regions are defined in table~\ref{tab:regionDefinitions}. Some background contributions are very small in specific control regions. \ratioDescription}
		\label{fig:sumPtCR}
	\end{center}
\end{figure}

Figure~\ref{fig:kinematicDistributionsCR} shows the good agreement obtained in kinematic distributions in the control regions. The \sumpt distribution for each control region is shown in figure~\ref{fig:sumPtCR}, which demonstrates good modelling of the background shape.

\clearpage
\subsection{Backgrounds from misidentified objects and non-prompt leptons}

The backgrounds from misidentified objects and non-prompt leptons are estimated with a data-driven matrix method, described in detail in ref.~\cite{atlas-topxsec}. In the electron channel, this contribution is dominated by the misidentification of hadronic jets, resulting in ``fake'' electrons. In order to make an estimate using this method, a sample enriched in multi-jet events is obtained by relaxing the electron selection criteria so as to increase the contribution from fakes. This is achieved by loosening the leading electron identification criteria from ``tight" to ``medium". The contribution from two or more fake electrons is found to be negligible. 

The numbers of data events in the sample which pass ($N_\mathrm{pass}$) and fail ($N_\mathrm{fail}$) the nominal tighter lepton selection requirements are counted. Defining $N_\mathrm{prompt}$ and $N_\mathrm{fake}$ as the numbers of events for which the leptons are prompt and fake, respectively, the following relationships hold:

\begin{equation}
N_\mathrm{pass} = \epsilon_\mathrm{prompt} N_\mathrm{prompt} + \epsilon_\mathrm{fake} N_\mathrm{fake},
\end{equation}
and
\begin{equation}
N_\mathrm{fail} = (1-\epsilon_\mathrm{prompt})N_\mathrm{prompt} + (1-\epsilon_\mathrm{fake}) N_\mathrm{fake}\, ,
\end{equation}
where $\epsilon_\mathrm{prompt}$ and $\epsilon_\mathrm{fake}$ are the relative efficiencies for prompt and fake leptons to pass the nominal selection, given that they satisfy the looser selection criteria. The simultaneous solution of these two equations gives a prediction for the number of events in data with fake leptons satisfying the nominal criteria, taken to be the estimated number of multi-jet events:

\begin{equation}
N_\mathrm{fake} = 
\frac{N_\mathrm{fail}-(1/\epsilon_\mathrm{prompt}-1)N_\mathrm{pass}}{1-\epsilon_\mathrm{fake}/\epsilon_\mathrm{prompt}} \> \> .
\label{eqn:multijet}
\end{equation}

The efficiencies $\epsilon_\mathrm{prompt}$ and $\epsilon_\mathrm{fake}$ are determined from control regions enriched in prompt-lepton or fake-lepton events, respectively. A fake-enhanced control sample is obtained starting from the preselection region, selecting events with exactly one lepton that satisfies the relaxed lepton criteria described above, $m_\mathrm{T} < 40\GeV$ and $m_\mathrm{T}+\met <60\GeV$. No \sumpt dependence in $\epsilon_\mathrm{fake}$ and $\epsilon_\mathrm{prompt}$ is observed, and the minimum  \sumpt requirement for these regions is set to $500\GeV$, compared with $\sumpt > 700$ GeV for the other control regions, in order to gain statistical power.

The efficiency for identifying fakes as prompt leptons is given by the fraction of the events in this control region that also pass the nominal lepton selection, after subtracting, in both instances, the estimated contribution from prompt-lepton backgrounds (derived from MC simulations, renormalised to match data in control regions, as described above).
For the electron channel, some dependence on the $\pt$ and $\eta$ of the leading electron is observed, which is taken into account by using $\pt$- and $\eta$-dependent $\epsilon_\mathrm{fake}$; they vary in the range $0.26$--$0.42$.

An equivalent procedure is followed for the muon channel, where the dominant contribution arises from non-prompt muons resulting from semileptonic decays, usually of heavy flavour. An event sample enhanced in these is formed by removing both the jet--muon overlap and muon isolation requirements.
For the muon channel, $\epsilon_\mathrm{non-prompt}$ is found to be negligibly small, consistent with zero:  $0.0043\pm 0.0040$~(stat), or $<0.011$ at $95\%$ CL; the resultant predicted background is negligible. 

The efficiency $\epsilon_\mathrm{prompt}$ is evaluated in a region with the same selection as the $Z$+jets control region, except for the relaxed \sumpt requirement, $500\GeV < \sumpt < 1500\GeV$, to match that used in the control region for fake and non-prompt leptons. The relative efficiency for identifying prompt leptons is obtained through the ratio of the number of events in which both leptons pass the nominal selection to those in which only one does. The measured values of $\epsilon_\mathrm{prompt}$ are $0.960\pm 0.007$ and $0.942\pm 0.007$ for electrons and muons, respectively, where the quoted uncertainties are statistical only.

\subsection{Background smoothing with fits}
\label{subSec:fittingStage}

\begin{figure*}[!htbp]
	\begin{center}
		\subfigure[]{ 
		\includegraphics[width=0.36\textwidth]{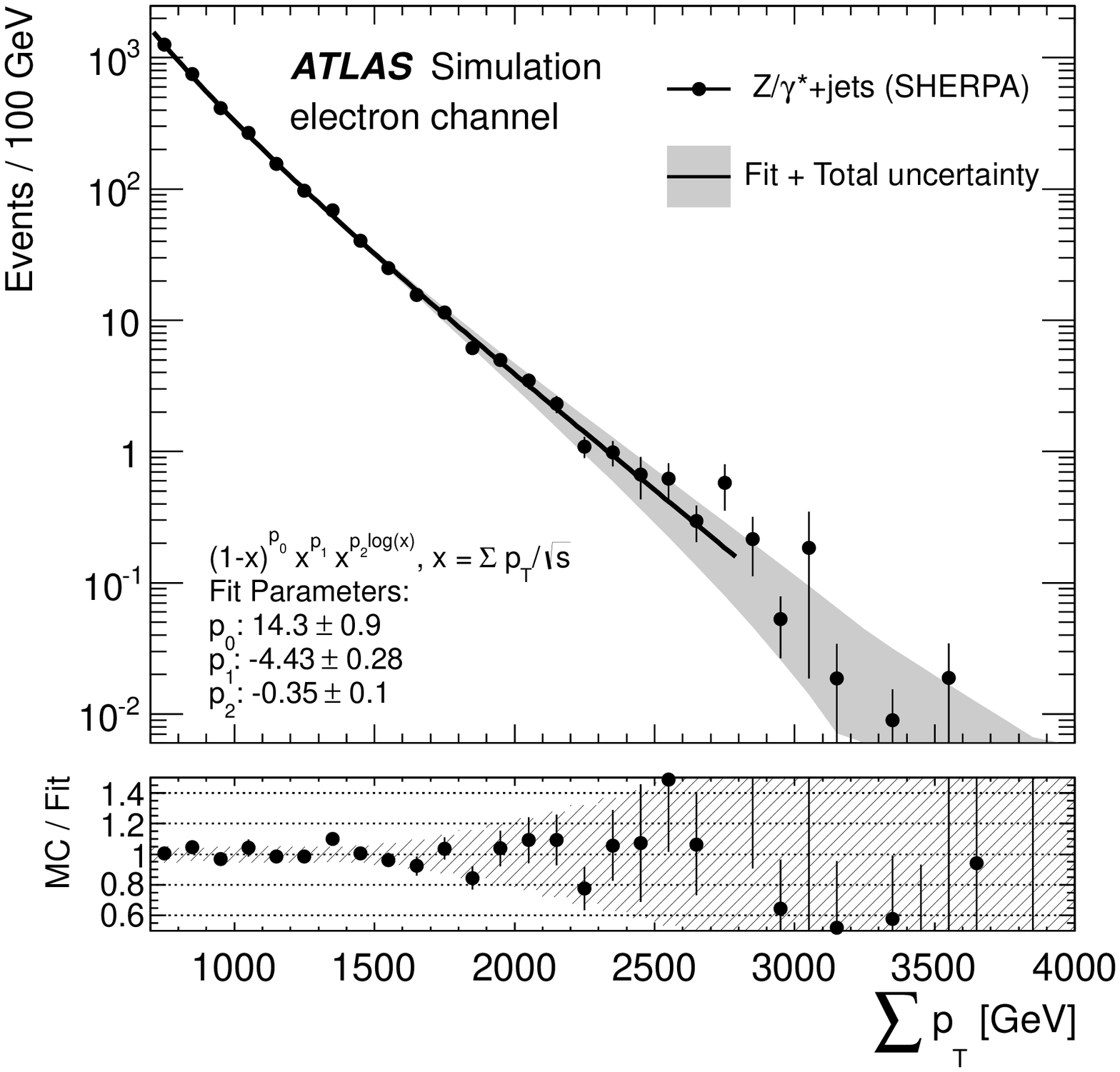}
		}
		\subfigure[]{
		\includegraphics[width=0.36\textwidth]{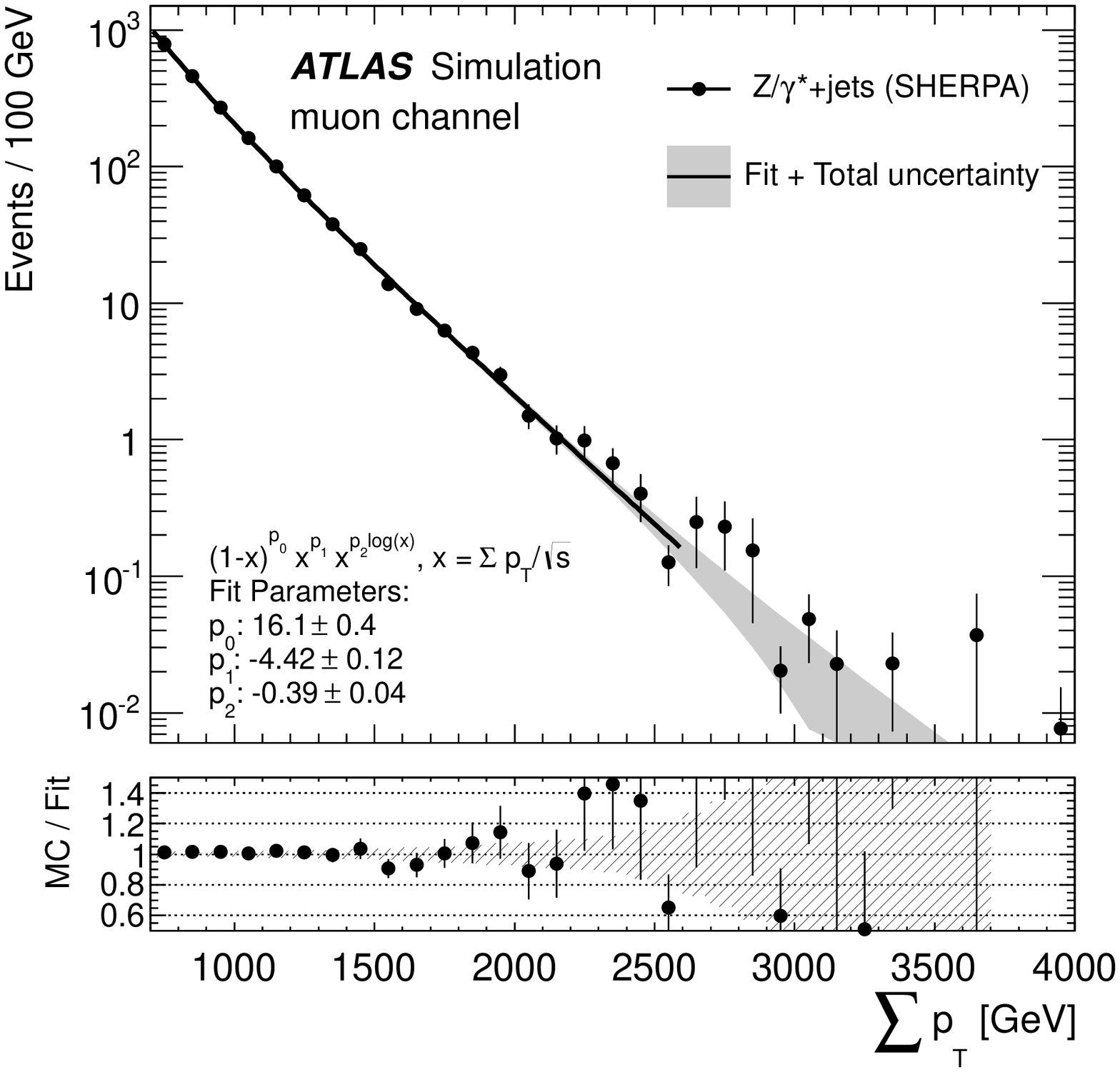}
		}
		\\
		\subfigure[]{ 
		\includegraphics[width=0.36\textwidth]{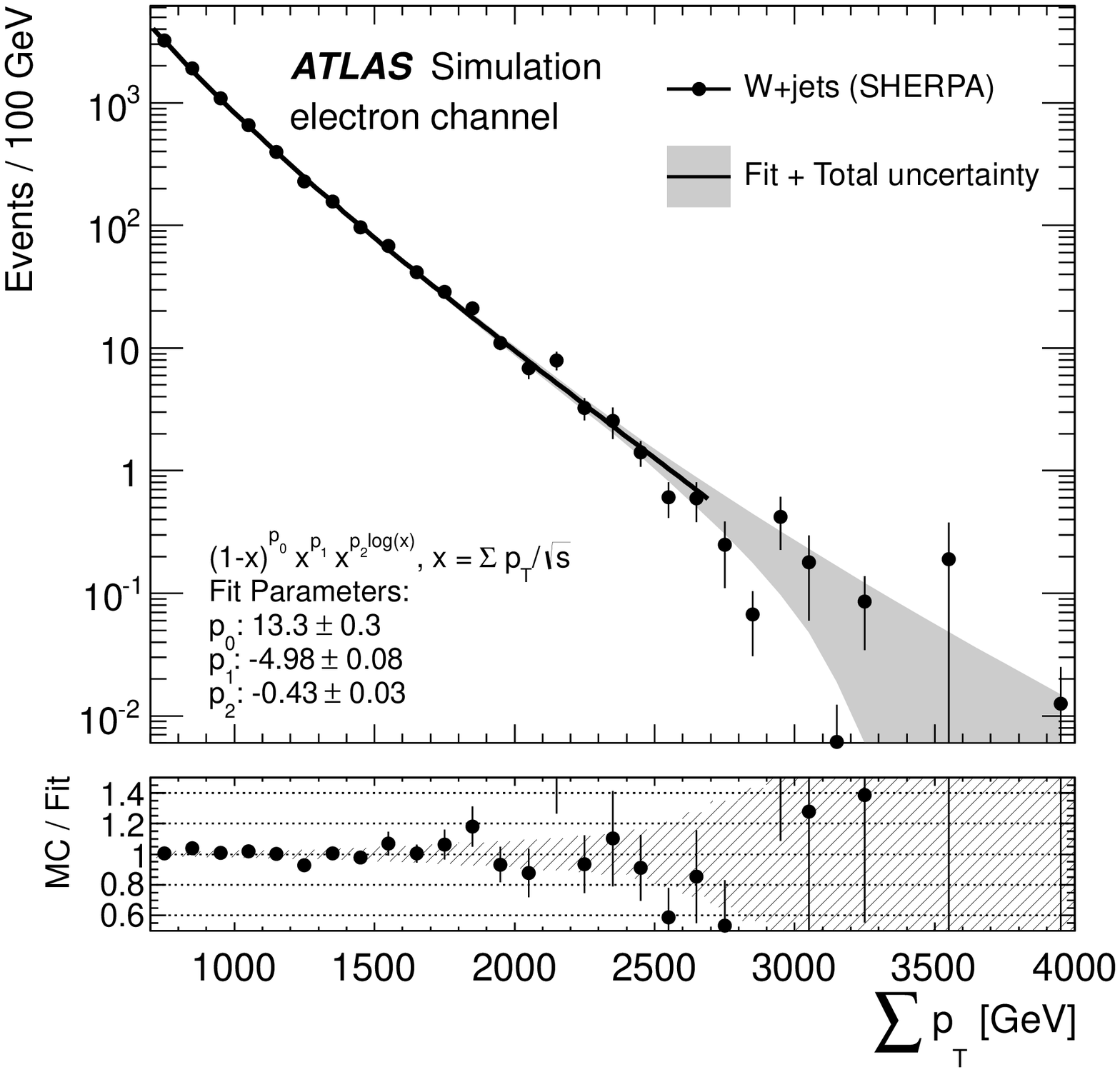}
		}
		\subfigure[]{ 
		\includegraphics[width=0.36\textwidth]{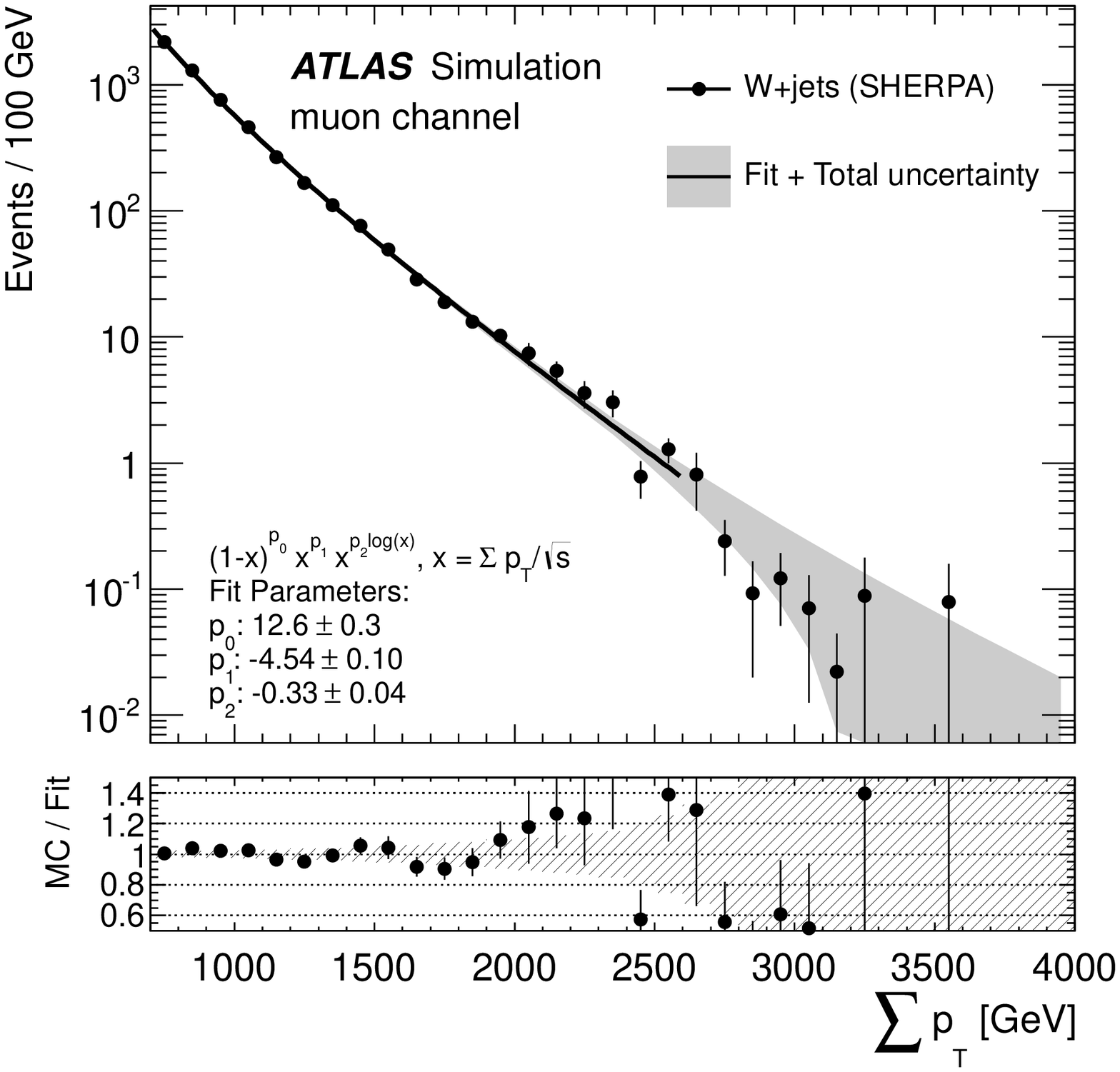}
		}
		\\
		
		\subfigure[]{ 
		\includegraphics[width=0.36\textwidth]{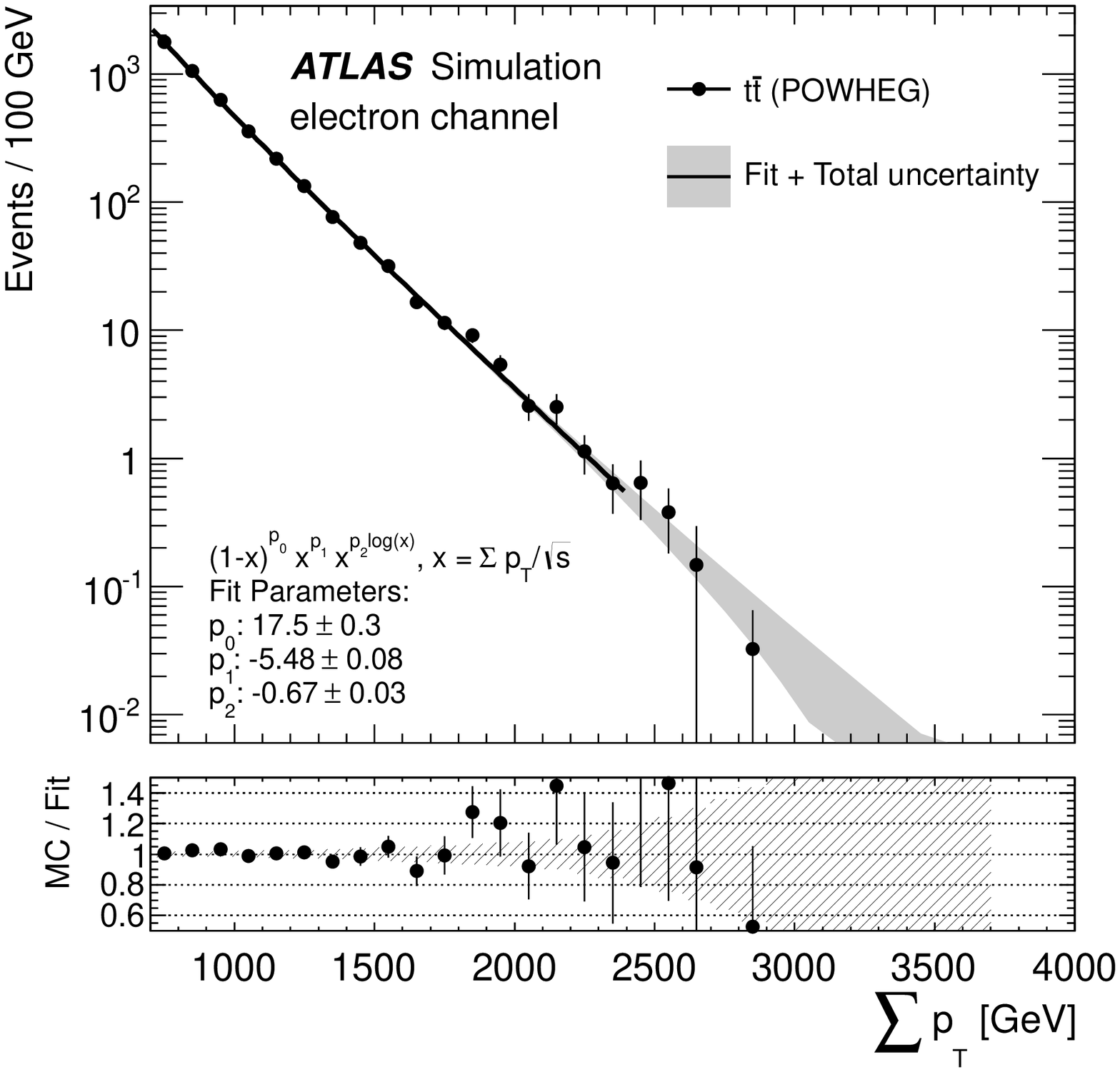}
		}
		\subfigure[]{
		\includegraphics[width=0.36\textwidth]{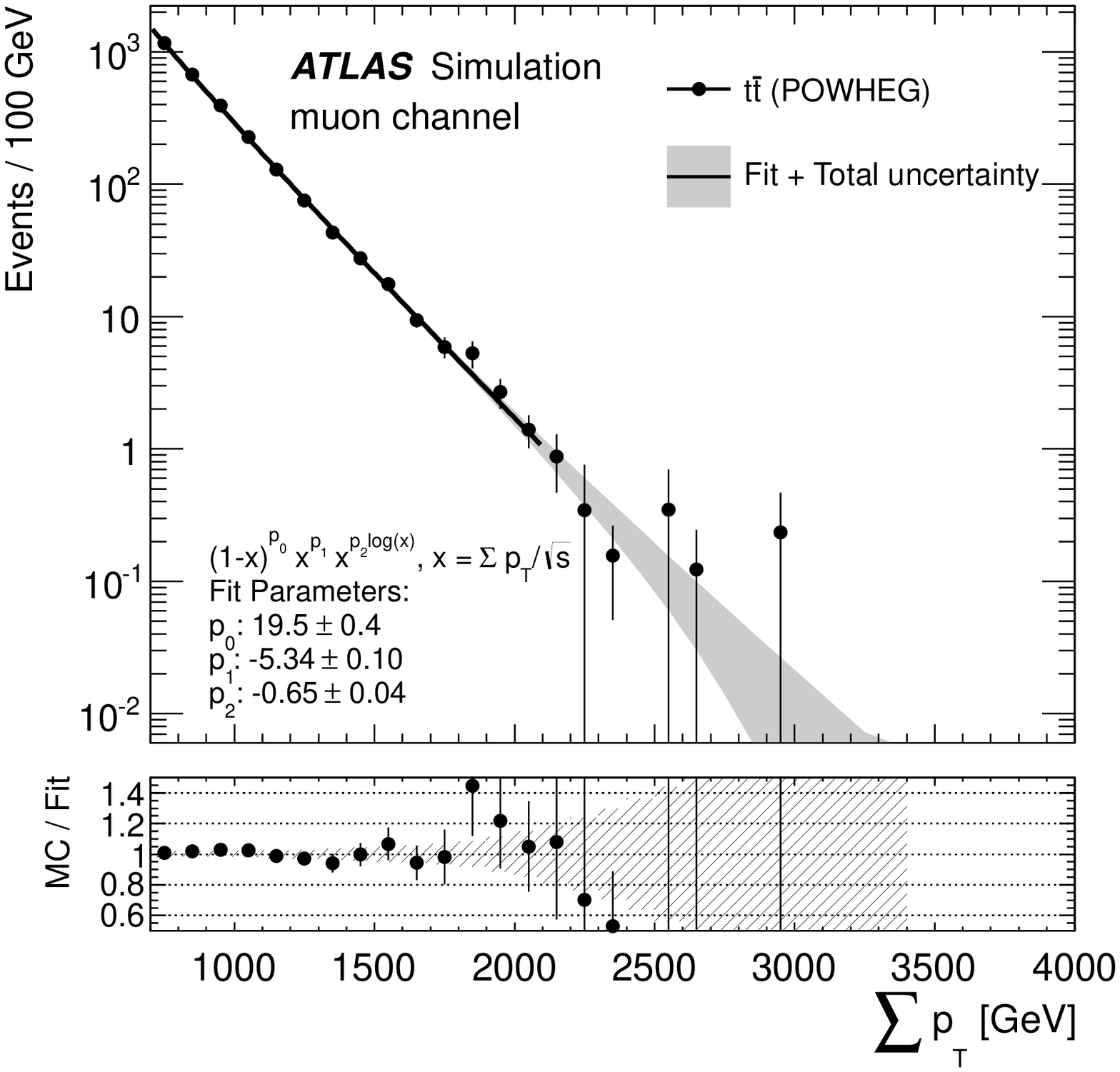}
		}
		\caption{The $\sumpt$ distributions and fit curves for (a, b) $Z+$jets; (c, d) $W+$jets ($+$ diboson); and (e, f) $t\bar{t}$ ($+$ single-top) MC-simulated events. Distributions for the electron channel are on the left while those for the muon channel are on the right. The shaded bands on the fit curves reflect the total uncertainty on the fit, including the systematic uncertainty discussed in section~\ref{subSec:fittingStage}. The length of the black line indicates the \sumpt range fitted. The lower panels show the ratio of the MC prediction to the fit, with the statistical uncertainty on the MC prediction (points), and separately, the fractional uncertainty on the fit (shaded band).}
		\label{fig:promptTemplates}
	\end{center}
\end{figure*}

\begin{figure}[!hbt]
\begin{center}
    \includegraphics[width=0.47\textwidth]{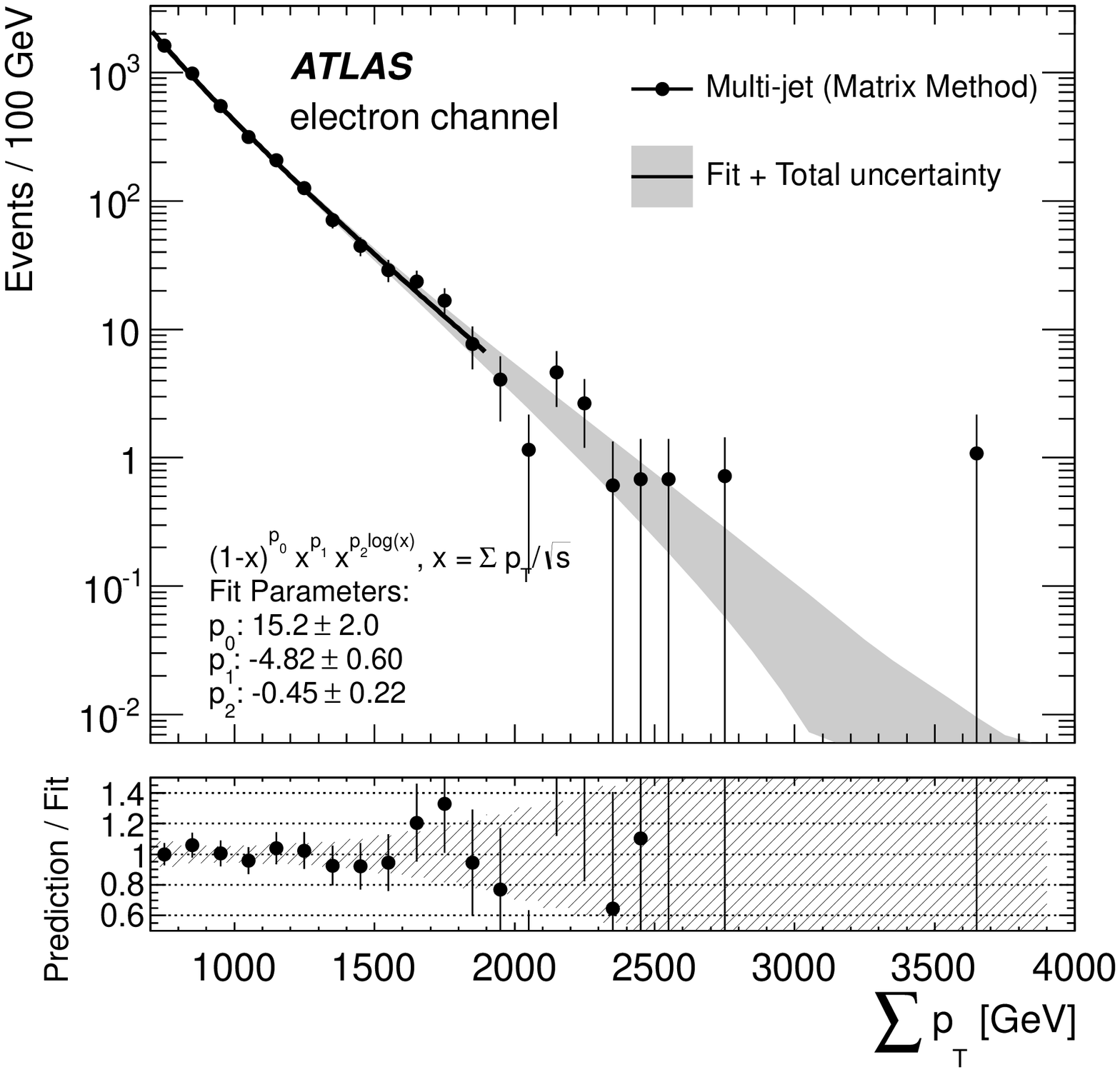}
\caption{The $\sumpt$ distribution and fit curve for the multi-jet background in the electron channel. The shaded band on the fit curve reflects the total uncertainty on the fit, including the systematic uncertainty discussed in section~\ref{subSec:fittingStage}. The length of the black line indicates the \sumpt range fitted. The lower panel shows the ratio of the matrix method's prediction to the fit, with the statistical uncertainty on the matrix method's prediction (points), and separately, the fractional uncertainty on the fit (shaded band).}
\label{fig:multijetTemplate}
\end{center}
\end{figure}

At high $\sumpt$, particularly beyond $\sumpt \approx 3500$~GeV, the numbers of events in the simulated background samples are small and consequently have large statistical uncertainties. To provide a more robust prediction in the signal region, the $\sumpt$ distributions of each individual background are fitted to an empirical function that enables the background shape to be smoothed and extended without being strongly affected by statistical fluctuations. This method reduces the statistical uncertainty, by using information at lower \sumpt to constrain the shape of the distribution, but introduces systematic uncertainties from the choice of binning and normalisation region, and from the choice of fit function. These are further discussed in section~\ref{sec:systuncert}. The fit function used is given by:

\begin{equation}
{\cal F}=\left(1-x\right)^{p_0}x^{p_1}x^{p_2\log \left(x\right)}\, ,
\end{equation}
where $x=\sumpt/\sqrt{s}$, and $p_0, p_1$ and $p_2$ are the parameters to be fitted. The overall normalisation is fixed by a combination of $p_0, p_1$ and $p_2$. The function was chosen for its stable and reliable description of the shape of the distributions over the full range of $\sumpt$. In previous studies~\cite{cdfDijets, ATLAS7TeVDiJets, ATLASResonancesDiLepton}, ATLAS and other experiments have found that this ansatz provides satisfactory fits to kinematic distributions.
The fit range begins at $\sumpt = 700$~GeV, and ends where the number of simulated events in a given bin is below five. The start- and end-points of the fit range, as well as the binning, are varied, and the results are consistent with the nominal fit within the statistical uncertainty. Although the default fits (shown in figures~\ref{fig:promptTemplates} and \ref{fig:multijetTemplate}) are of high quality and stability, there is an uncertainty associated with the choice of background fit function. To assess this, an alternative function was chosen that succeeds in describing the distributions at low and intermediate $\sumpt$ but has a different shape than the nominal function at high $\sumpt$, where the numbers of simulated events are smaller. 
This function is given by:

\begin{equation}
{\cal F}_\mathrm{alt}=\frac{p_0}{x} \left( 1-p_1 x\right)^{p_2}\, ,
\label{eqn:fitfunc}
\end{equation}
where $x=\sumpt/\sqrt{s}$, and $p_0, p_1$ and $p_2$ are the parameters to be fitted.

In the bins where the prediction from this alternative function falls outside the nominal fit uncertainty, the difference between the nominal and alternative functions is used as the fit uncertainty, i.e. an envelope of them is taken and symmetrised, to be conservative, ensuring that the total fit uncertainty covers alternative functions and the inherent uncertainty of the fit itself. 

The $\sumpt$ distributions for each MC-simulated prompt-lepton background are displayed in figure~\ref{fig:promptTemplates}; the curves shown represent the results of the binned maximum likelihood fits. The multi-jet background in the electron channel estimated from data is fitted to the same function, as shown in figure~\ref{fig:multijetTemplate}. The fit quality is high, with typical $\chi^2$/d.o.f. values between 0.9 and 1.6.

The fitted shapes of the individual backgrounds are combined according to their relative predicted contributions (as discussed in the preceding sections, computed in a subset of the sideband region, $1000 < \sumpt < 1500$~GeV) to give an overall background template shape. 
In order to reduce the systematic uncertainty, this is normalised to data in this same region by a minimisation of the $\chi^2$ difference between the data and the background template. This results in a normalisation consistent with that determined from the control regions within the $4\%$ uncertainty resulting from the statistical uncertainty on the data in these bins. The resulting background estimate gives a smooth and stable prediction at all values of \sumpt.

\FloatBarrier

\section{Systematic uncertainties}
\label{sec:systuncert}

Sources of systematic uncertainty in the background prediction are taken into account.
These are reduced by the normalisation to data in the control regions, making the analysis insensitive to \sumpt-independent uncertainties, such as those on the luminosity measurement (this uncertainty is applied to the signal expectation). Uncertainties on the shape of the $\sumpt$ distribution, on the other hand, can have an impact on the background prediction. 

The uncertainty from the fit to the backgrounds is the dominant systematic uncertainty. Its impact on the background yield varies from $25\%$ ($20\%$) for the $\sumpt>2000$~GeV region, to $140\%$ ($190\%$) for the $\sumpt>3200$~GeV region, for the electron (muon) channel. The systematic uncertainties resulting from variations of the fit range and alternative choices for the $\sumpt$ region used to normalise the background template are found to be negligible.

The experimental uncertainties are small compared to the fit uncertainty in all signal regions considered. Their impact is assessed by applying each systematic uncertainty to the background samples, changing both the relative fractions of the backgrounds and their shapes. This is then propagated to the fits, and a new spectrum is obtained. The difference between the nominal prediction and the new prediction determines the systematic uncertainty. The most important experimental systematic uncertainty comes from the jet energy scale. This is determined using in situ techniques~\cite{Aad:2011he}, and gives rise to systematic uncertainties of $2$--$10\%$ for the lower $\sumpt$ signal regions, and no more than $20\%$ for the highest $\sumpt$ signal regions. Systematic uncertainties from jet energy resolution and $b$-tagging~\cite{btagging,btagging2} are found to be small ($< 5$\%), even for the highest \sumpt thresholds considered, while uncertainties from missing transverse energy, and lepton scale, identification and resolution are found to be completely negligible. Additional uncertainties arise from the choice of MC generators ($5$--$10\%$, comparing the nominal generators for the three main prompt backgrounds with {\sc Alpgen}) and limited knowledge about the parton distribution functions at high $\sumpt$ ($2$--$10\%$). The latter includes both the appropriate PDF error set ({\sc CT10} for all major backgrounds) and variations in choice of PDF through comparison with {\sc MSTW2008nlo}. Uncertainties from the choice of hadronisation and factorisation scales were considered and found to be negligible.

Uncertainties on the signal yields include the detector response uncertainties discussed above, luminosity and statistical uncertainties. Uncertainties that affect both the background and the signal are taken to be completely correlated. In addition, a $5\%$ systematic uncertainty is included on the signal normalisation, corresponding to the maximum observed acceptance difference between MC-simulated samples using full \geant simulation and the fast simulation. The effect of PDF variations on the signal acceptance is found to be negligible. The theoretical and modelling uncertainties on these states motivate the choice of benchmark models and are discussed in section~\ref{sec:mcsamples}; limits are set for exactly these benchmark models.

\section{Results and interpretation}

\begin{figure*}[!hbp]
\begin{center}
  \subfigure[]{ 
    \includegraphics[width=0.6\textwidth]{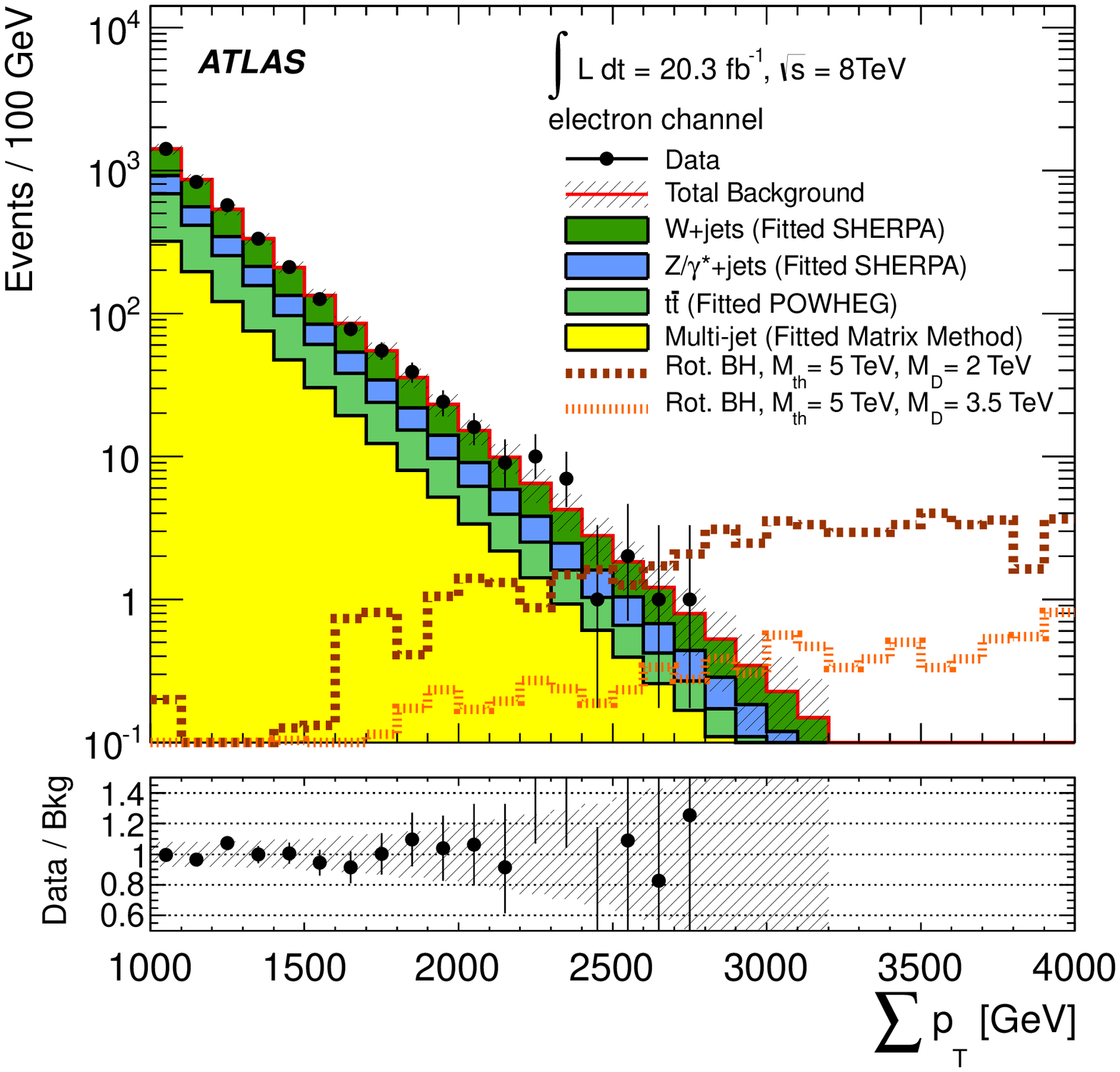}
  }
  \subfigure[]{ 
    \includegraphics[width=0.6\textwidth]{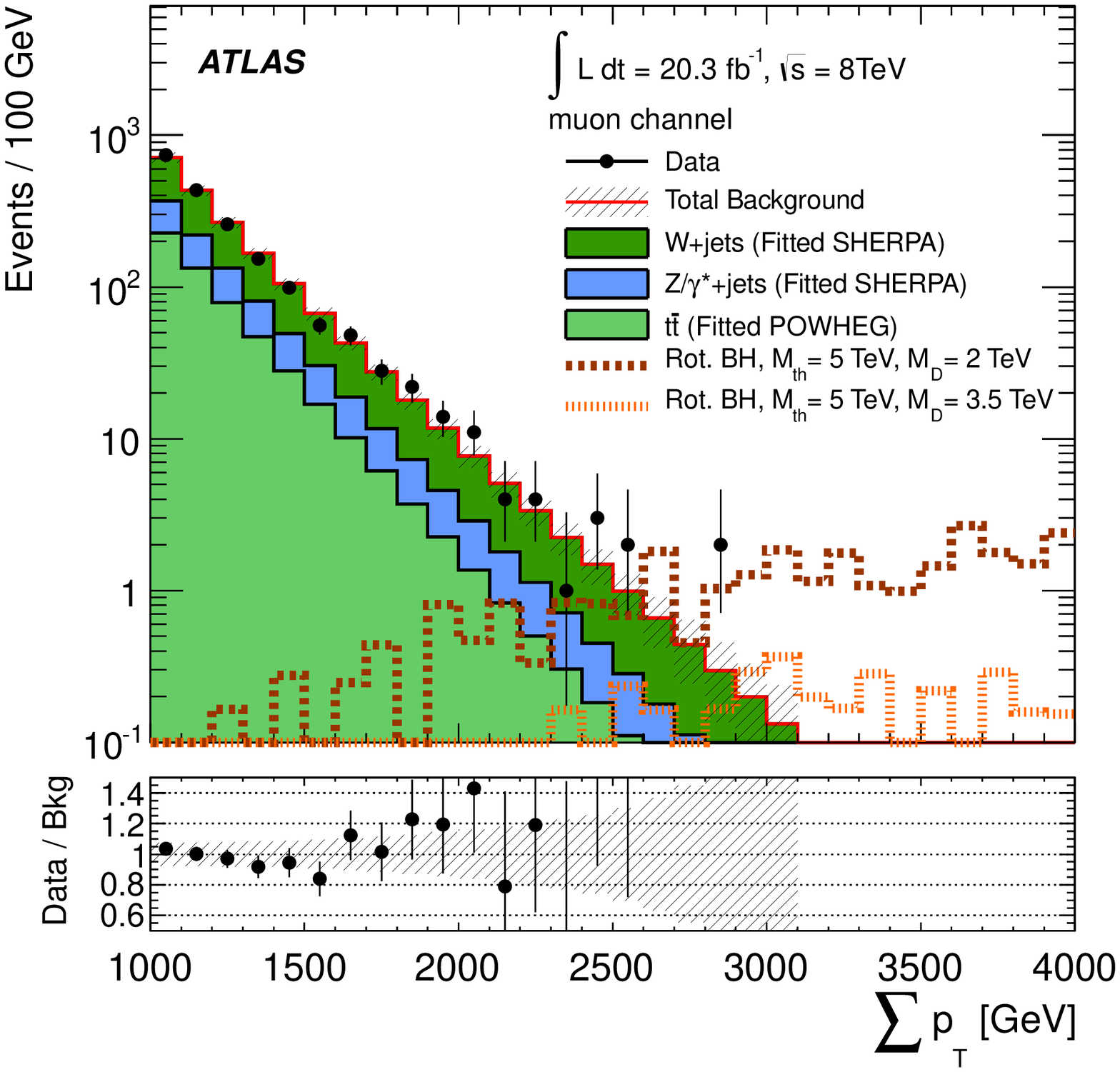}
  }
\caption{The  $\sumpt$ distributions in the (a) electron and (b) muon channels. Two representative signal distributions for rotating black holes with $n =6$ are overlaid to illustrate the signal properties. \ratioDescription}
\label{fig:finalsumpt}
\end{center}
\end{figure*}

The $\sumpt$ distributions observed from data and predicted from SM processes for the electron and muon channels in the sideband and signal regions are given in figure~\ref{fig:finalsumpt}, with two representative signal distributions superimposed: rotating black holes with $n = 6$ and \MTH = 5 TeV, one with \MD = 2 TeV, the other with \MD = 3.5 TeV.\footnote{These points have cross-sections of 46.3 pb and 12.9 pb, respectively, as calculated by \charybdis.} The yields in the signal region for various choices of $\sumpt$ threshold are shown in table~\ref{tab:results}. In both channels, the fraction of the $W+$jets background, already dominant at lower \sumpt, increases further for higher \sumpt. The $W+$jets background constitutes $45\%$ ($66\%$) of the background yield above $\sumpt = 2000$~GeV in the electron (muon) channel; the contribution from $Z+$jets is $19\%$ ($17\%$), whereas $t\bar{t}$ accounts for $15\%$ ($17\%$), with the remainder in the electron channel being multi-jet events. For both channels, no data events are observed above \sumpt = 3000 GeV, in agreement with the background estimate.

\begin{table*}[!htb]
  \begin{center}
   \begin{tabular}{|c||c|c||c|c|}\hline
     & \multicolumn{2}{c||}{Electron Channel} & \multicolumn{2}{c|}{Muon Channel} \\ Min. $\sumpt$ [GeV] & Expected Background & Data & Expected Background & Data\\ \hline 
        $2000$  & $44\pm 12$ & $47$ & $22.8\pm 5.4$ & $27$\\  
        $2200$  & $19\pm 7$ & $22$ & $10.1\pm 3.2$ & $12$\\  
        $2400$  & $8.2\pm 3.7$ & $5$ & $4.5\pm 1.9$ & $7$ \\  
        $2600$  & $3.5\pm 2.1$ & $2$ & $2.0\pm 1.3$ & $2$ \\  
        $2800$  & $1.5\pm 1.2$ & $0$ & $0.89\pm 0.82$ & $2$ \\  
        $3000$  & $0.65 ^{+0.69} _{-0.65}$ & $0$ & $0.40 ^{+0.53} _{-0.40}$ & $0$ \\  
        $3200$  & $0.28 ^{+0.40} _{-0.28}$ & $0$ & $0.18 ^{+0.34} _{-0.18}$ & $0$ \\  
                  \hline      \end{tabular}
    \caption{Expected SM background and observed event yields for the electron and muon channels, for the signal regions of this search. The quoted uncertainties on the background yields are the combined statistical and systematic uncertainties.}
    \label{tab:results}
    \end{center}
\end{table*}

No significant excess is observed beyond the Standard Model expectation for all choices of $\sumpt$ threshold: $p$-values for the background-only hypothesis are in the range 0.2 - 0.5.~\footnote{The $p$-value is truncated at 0.5, since only upward fluctuations of the background are taken into account.} Consequently, limits are set on the fiducial cross-section and on TeV-scale gravity benchmark models, using the modified frequentist CL$_{\mathrm s}$ prescription described in ref.~\cite{Read:2002hq}. It compares the number of observed events in data with the SM expectation, using the profile likelihood ratio as test statistic. All systematic uncertainties and their correlations are taken into account via nuisance parameters.

\clearpage

\begin{figure*}[!hp]
	\centering
	\includegraphics[width=0.6\textwidth]{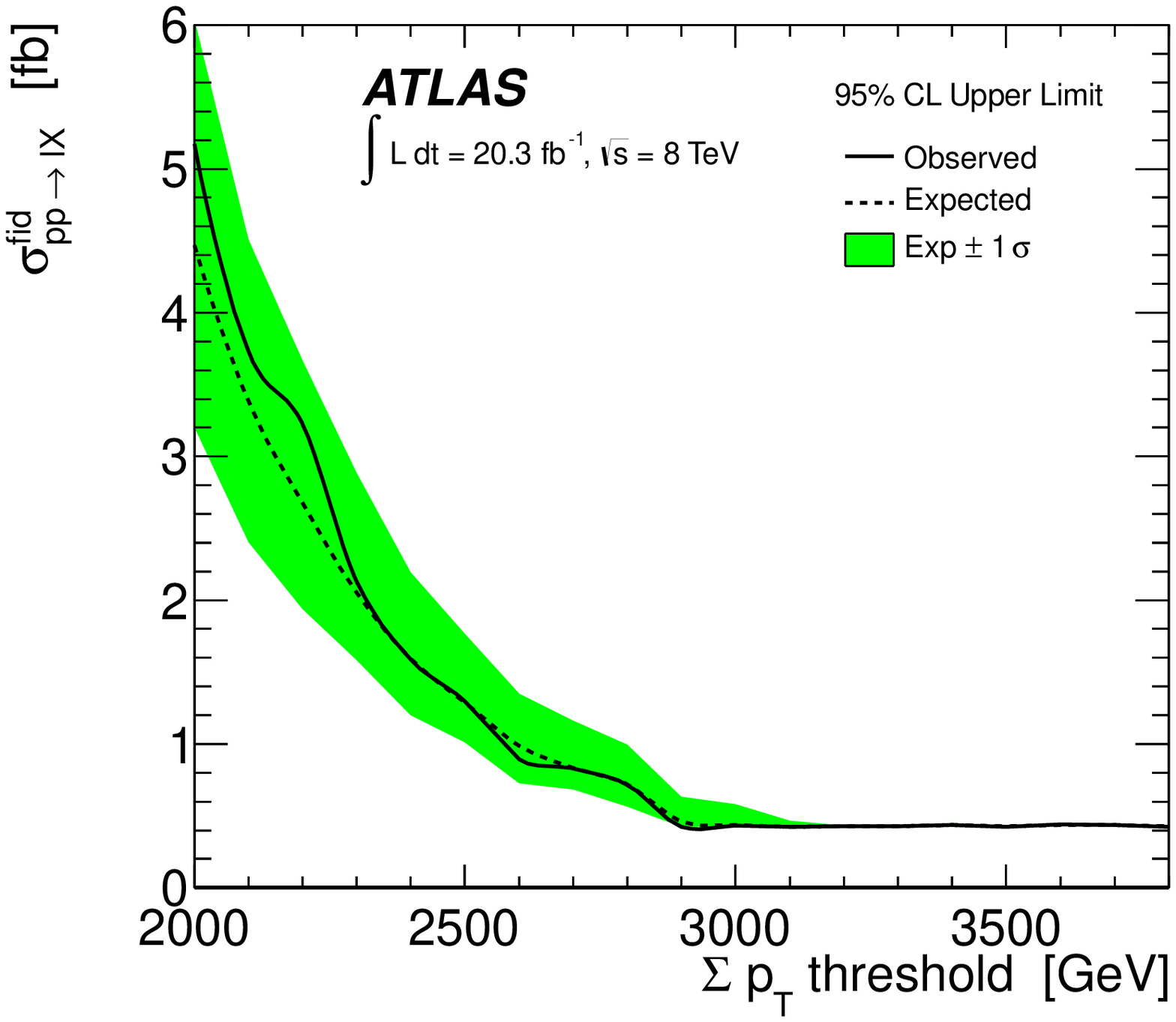}
	\caption{Upper limits on the fiducial cross-sections $\sigma^\mathrm{fid}_{pp\to \ell X}$ for the production of final states with at least three objects passing a 100 GeV $\pt$ requirement including at least one lepton, and $\sumpt$ above threshold, for all final states with at least one electron or muon. The observed and expected 95\% CL limits according to the CL$_{\mathrm s}$ prescription are shown, as well as the $\pm1\sigma$ %and 2$\sigma$ 
bounds on the expected limit. \label{fig:limits_combined_fidcs}
	}
\end{figure*}

\begin{figure*}[!hp]
	\subfigure[]{
	\includegraphics[width=0.5\textwidth]{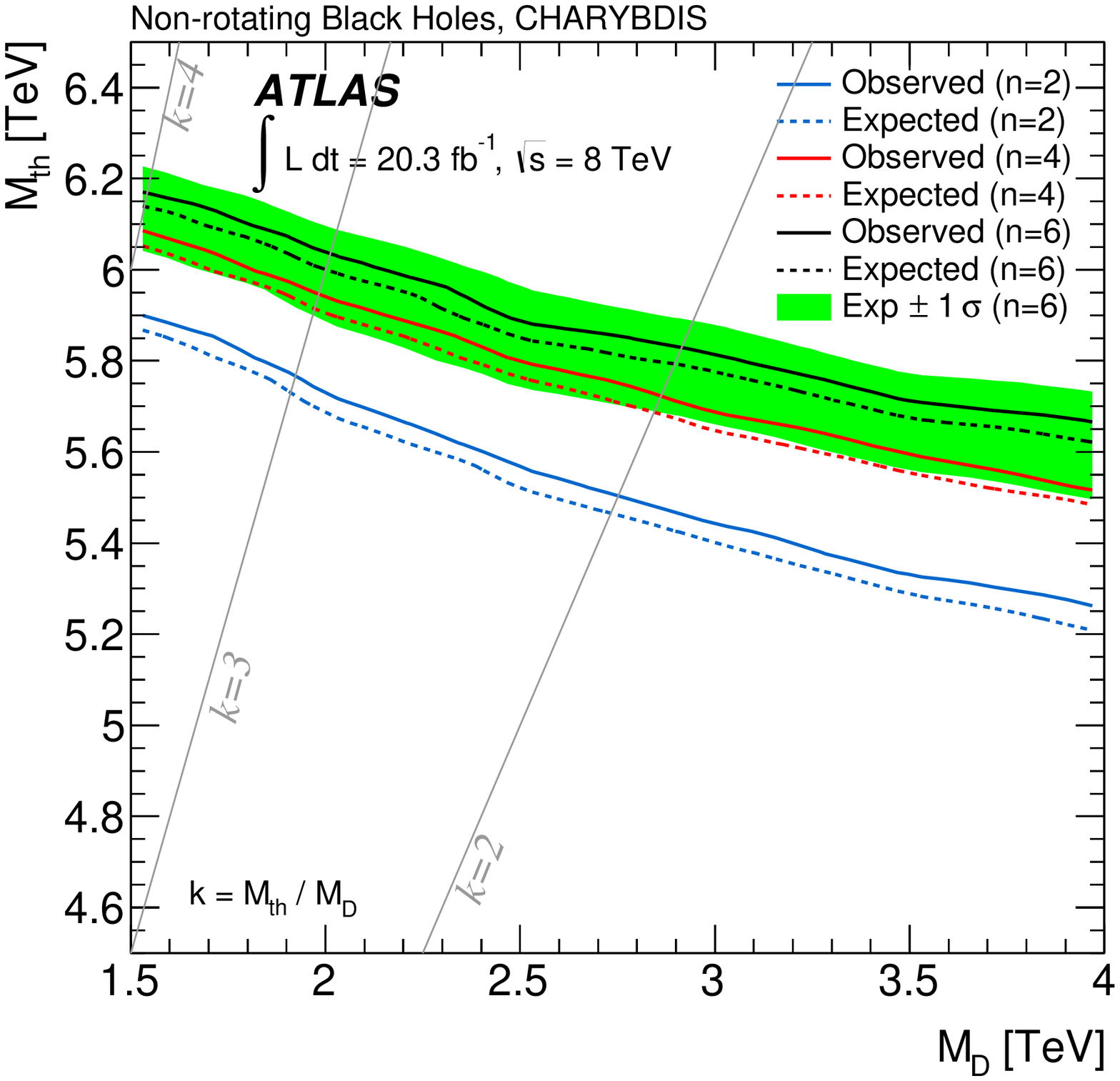}
	}
	\subfigure[]{
	\includegraphics[width=0.5\textwidth]{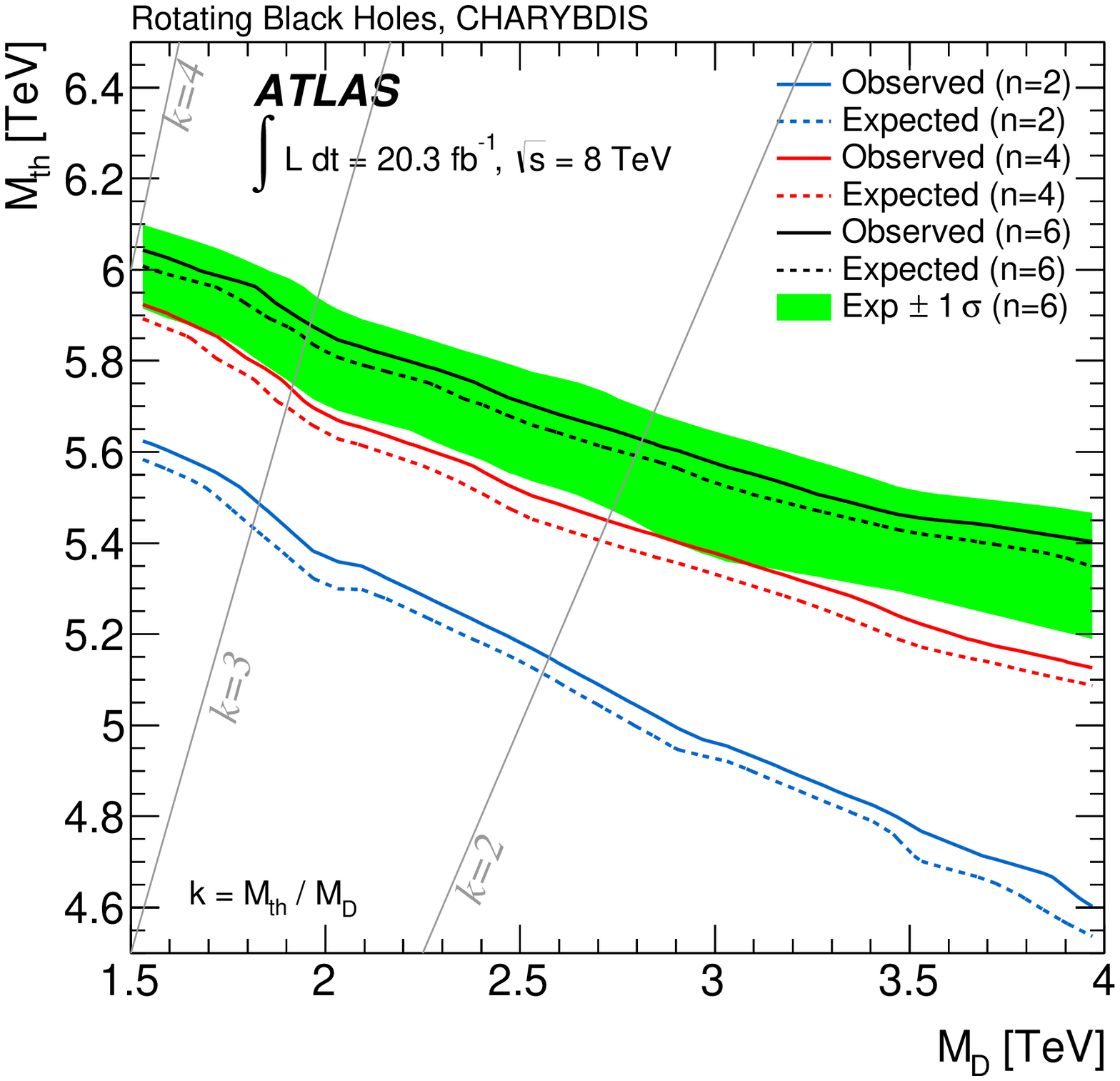}
	}
	\subfigure[]{
	\includegraphics[width=0.5\textwidth]{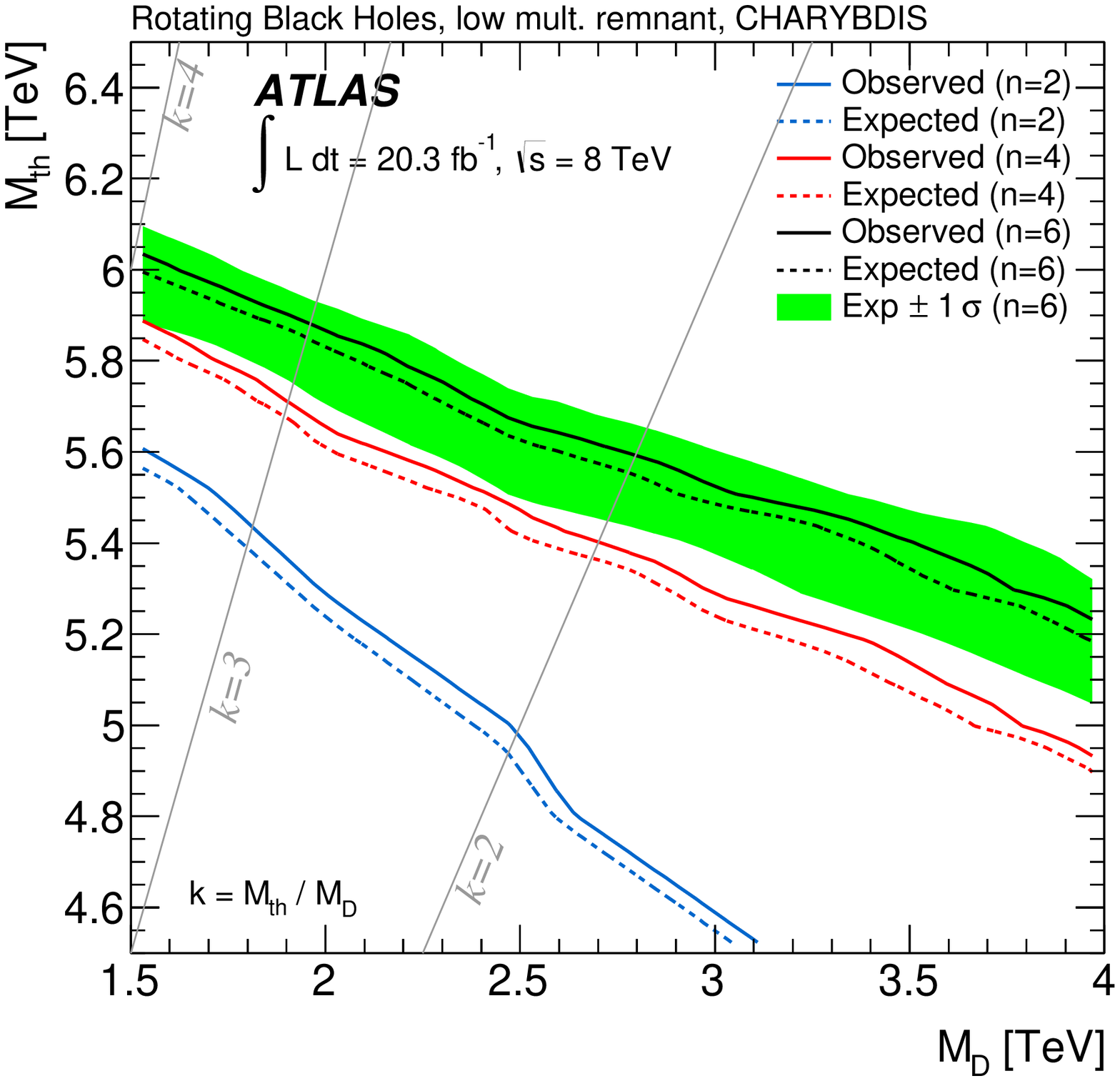}
	}
	\subfigure[]{
	\includegraphics[width=0.5\textwidth]{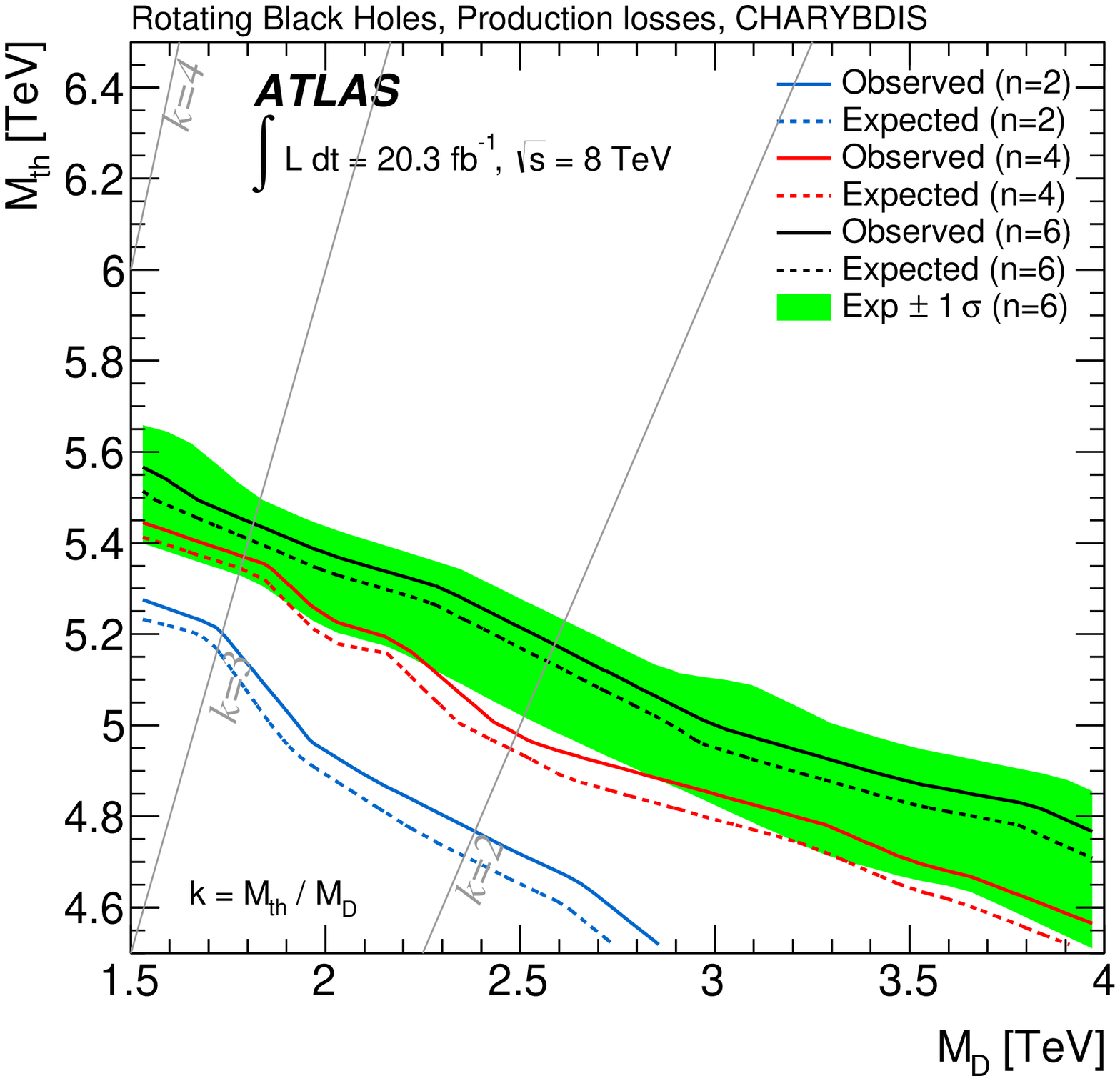}
	}
	\caption{\label{fig:limits_combined_Charybdis2}
	The exclusion limits in the $\MTH$--$\MD$ plane, with electron and muon channels combined, for (a) non-rotating and (b) rotating black hole models in two, four and six extra dimensions, simulated with \charybdis. The lower panes show limits for (c)  rotating black holes with low multiplicity remnant decays and (d) with production phase losses turned on.
	The solid (dashed) lines show the 
	observed (expected) $95\%$ CL limits, with the shaded band illustrating the expected $\pm~1~\sigma$ variation of the $n=6$ expected limits. The $\pm1\sigma$ variation is comparable for the $n=2$ and $n=4$ models. 
	Masses below the corresponding lines are excluded. The lighter grey lines indicate constant $k=\MTH/\MD$.
	}
\end{figure*}
\begin{figure*}[!htbp]

	\begin{center}
	\includegraphics[width=0.5\textwidth]{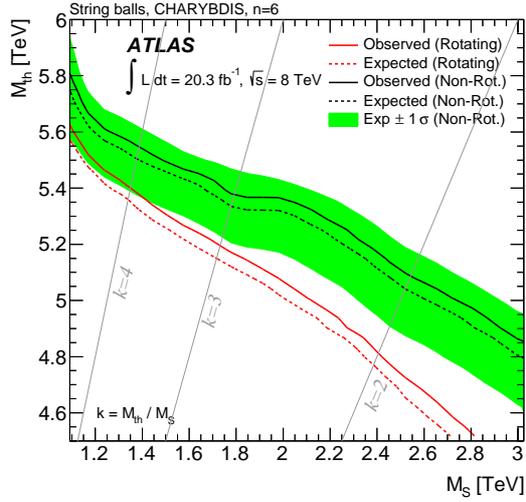}
	\caption{\label{fig:limits_combined_stringballs}
	The exclusion limits in the $\MTH$--$\MS$ plane, with electron and muon channels combined, for rotating and non-rotating string ball models with six extra dimensions.
	The solid (dashed) lines show the 
	observed (expected) $95\%$ CL limits, with the band illustrating the expected $\pm~1~\sigma$ variation of 
	the expected limits for the non-rotating case. The $\pm1\sigma$ variation is comparable for the rotating case. Masses below the corresponding lines are excluded. The lighter grey lines indicate constant $k=\MTH/\MS$. 
	}
	\end{center}
\end{figure*}
\begin{figure*}[!hp]
	\subfigure[]{
	\includegraphics[width=0.5\textwidth]{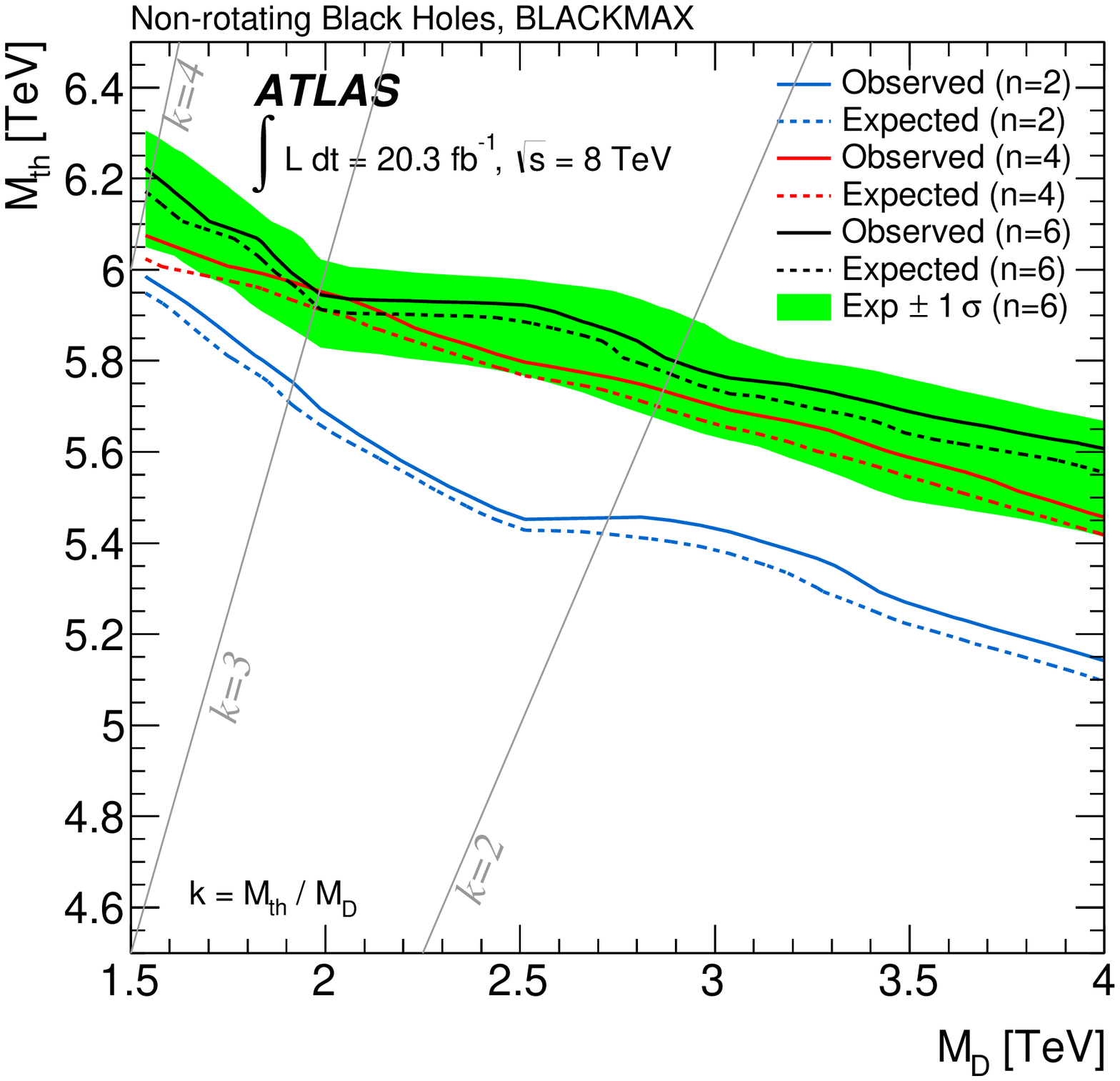}
	}
	\subfigure[]{
	\includegraphics[width=0.5\textwidth]{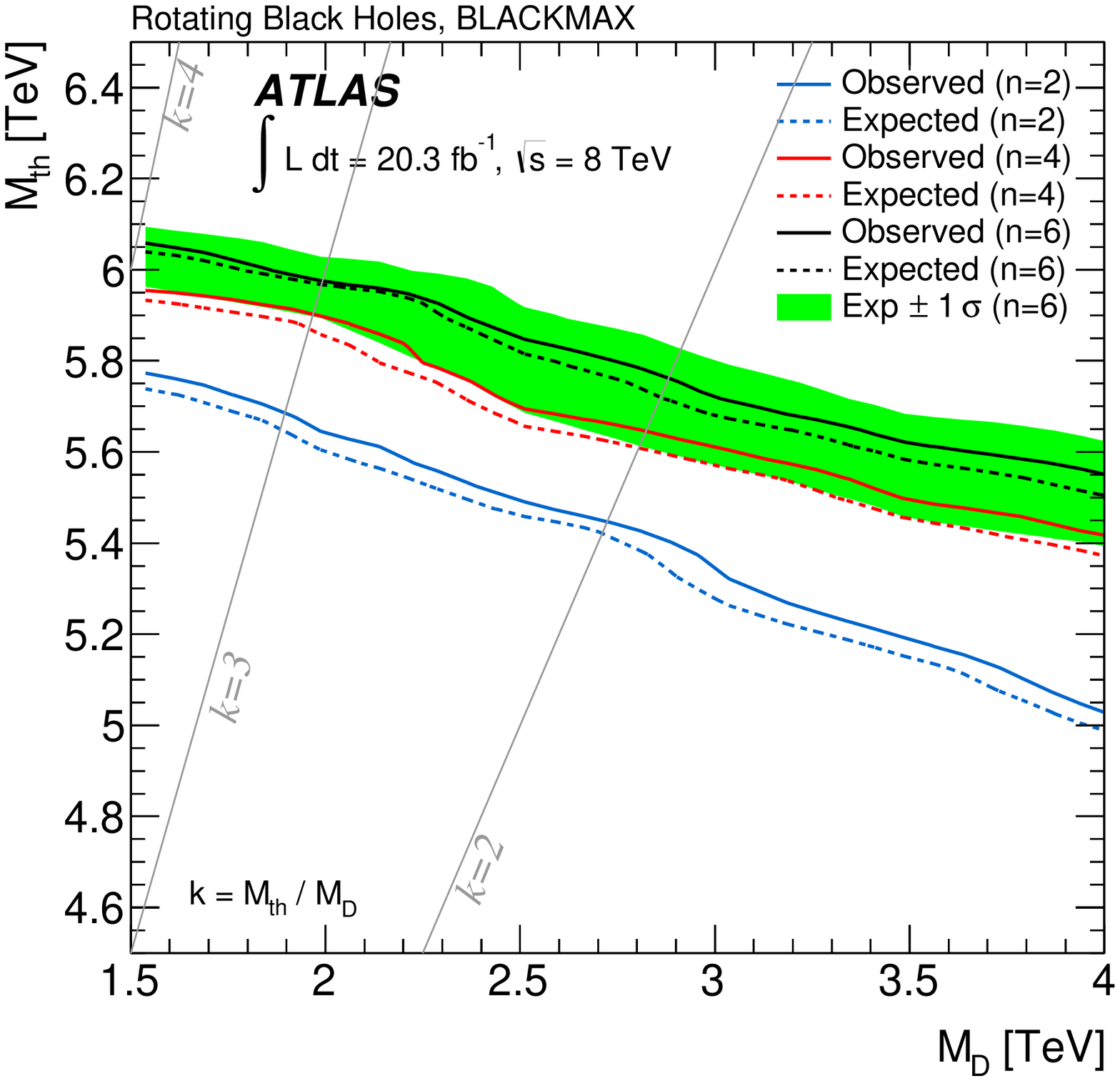}
	}
	\subfigure[]{
	\includegraphics[width=0.5\textwidth]{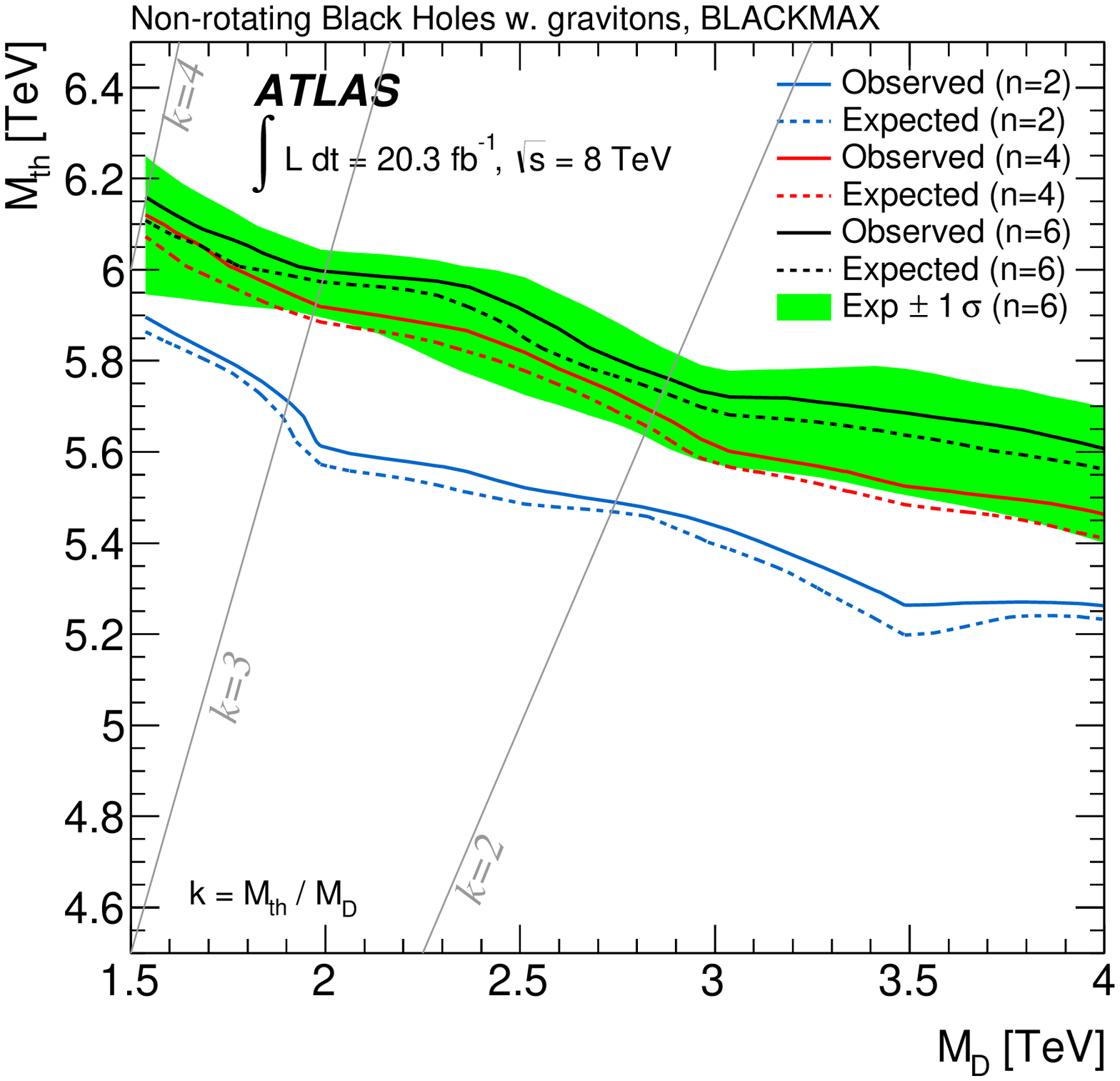}
	}
	\subfigure[]{
	\includegraphics[width=0.5\textwidth]{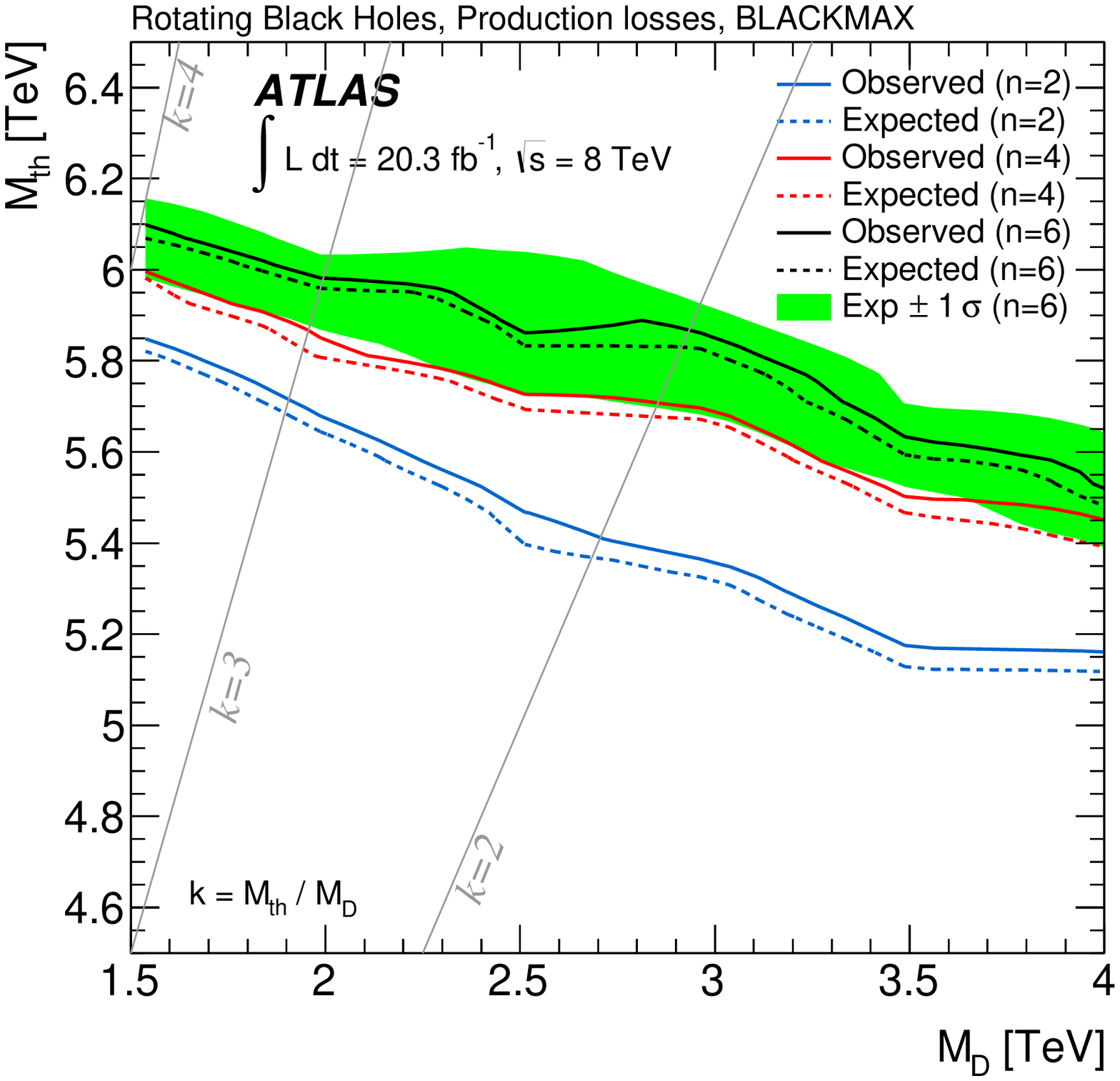}
	}
	\caption{\label{fig:limits_combined_BlackMax}
	The exclusion limits in the $\MTH$--$\MD$ plane, with electron and muon channels combined, for (a) non-rotating and (b) rotating black hole models two, four and six extra dimensions, simulated with \blackmax.
	The lower panes show limits for (c) non-rotating black holes with graviton emission and (d) rotating black holes with production phase losses turned on.
	The solid (dashed) lines show the 
	observed (expected) $95\%$ CL limits, with the shaded band illustrating the expected $\pm~1~\sigma$ variations of the $n=6$ expected limits. The $\pm1\sigma$ variation is comparable for the $n=2$ and $n=4$ models. %and $2\sigma$ variations of the expected limits. 
	Masses below the corresponding lines are excluded. The lighter grey lines indicate constant $k=\MTH/\MD$. 
	}
\end{figure*}

\begin{figure*}[!hp]
	\begin{center}
		\includegraphics[width=0.5\textwidth]{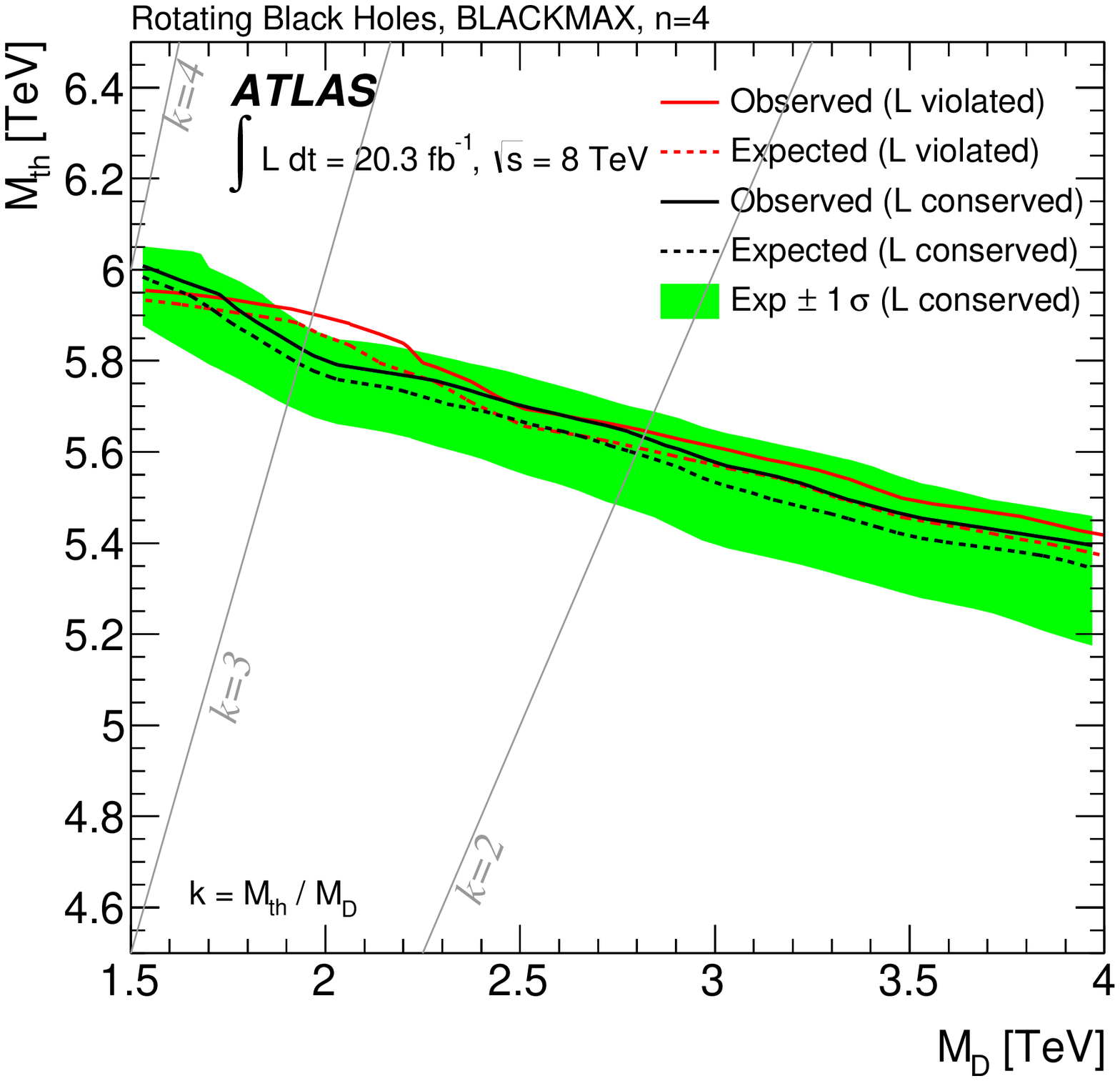}
		\caption{\label{fig:limits_combined_c1}
		The exclusion limits in the $\MTH$--$\MD$ plane, with electron and muon channels combined, for a rotating black hole models with lepton number conservation or violation in four extra dimensions.
		The solid (dashed) lines show the 
		observed (expected) $95\%$ CL limits, with the band illustrating the expected $\pm~1~\sigma$ variation of
		the expected limits for the model with lepton number conservation. The $\pm1\sigma$ variation is comparable for the lepton violating model. The lighter grey lines indicate constant $k=\MTH/\MD$
		}
	\end{center}
\end{figure*}

Limits on the fiducial cross-section $\sigma^\mathrm{fid}_{pp\to \ell X}$, defined as that part of the total cross-section which is within the kinematic limits of the measurement, are calculated at 95\% CL. This requires the determination of a reconstruction efficiency factor, $C_{pp\to \ell X}$, that converts the observed signal yield ($N_\mathrm{signal}$) to the yield in the fiducial region at the generator level:

\begin{equation}
\sigma^\mathrm{fid}_{pp\to \ell X} = \frac{N_\mathrm{signal}}{{\cal L}\cdot C_{pp\to \ell X}}\, ,
\end{equation} 
where ${\cal L}$ is the integrated luminosity used in the analysis. 

The fiducial regions at generator level\footnote{This includes parton showering and jet clustering, using the anti-$k_t$ algorithm with $R$ = 0.4 on stable particles.} for the electron and muon channels are defined from the simulated \charybdis signal events with final states that pass the following requirements:
the leading lepton is a prompt electron or muon\footnote{Electrons and muons originating from $\tau$ leptons, heavy gauge bosons or directly from the black hole are considered to be prompt.} within the experimental acceptance, with $\pt > 100\GeV$ and separated from jets with $\pt>60$~GeV by $\Delta R$(lepton, jet)~$>0.4$; there are at least two additional jets or leptons with $\pt > 100\GeV$ present in the event, and $\sumpt$ is above the signal region threshold. The requirements are summarised in table~\ref{tab:truthAnalysisDefinition}. 
Additionally, given the appropriate $C_{pp\to \ell X}$, the channels can be combined to set a limit on  $\sigma^\mathrm{fid}_{pp\to \ell X}$ for anomalous production of final states with a lepton ($e$ or $\mu$). In this ``lepton'' channel limit, the expectations of the two channels are summed, with the uncertainties combined, taking their correlations into account. 
For the wide range of models considered (black hole states, string ball states, rotating and non-rotating, low- and high-multiplicity remnant states, etc.), $C_{pp\to \ell X}$ varies from  $47$\% to $82\%$ ($22$\% to $46\%$) for the electron (muon) channel, and $34$--$64\%$ for the combined case. It is lowest for two extra dimensions, due to the lower multiplicity of the final state. The fiducial cross-section limits are given in table~\ref{tab:fidexclusion}, where the lowest overall values of $C_{pp\to \ell X}$ within the range of the model dependence are used (which correspond to $n = 2$). Figure~\ref{fig:limits_combined_fidcs} shows the limits on $\sigma^\mathrm{fid}_{pp\to \ell X}$ for the combined channels using this most conservative value of $C_{pp\to \ell X}$.

\begin{table*}
	%\footnotesize
	\begin{center}
		\begin{tabular}{|c|c|}%{\textwidth}@{\extracolsep{\fill}
			\hline
			\multicolumn{2}{|c|}{\textbf{Object Selection}} \\
			\hline
			\textbf{Object}                                & \textbf{Requirement} \\
			\hline 
			Electrons      &  $\pt > 60$~GeV, $1.37 < |\eta|$ or $1.52 < |\eta| < 2.47$ \\ \hline
			Muons      &   $\pt > 60$~GeV, $|\eta| < 2.4$  \\ \hline
			Jets      &  anti-$k_t$ truth jet, $R = 0.4$ \\
			&  $\pt > 60$~GeV, $|\eta| < 2.8$  \\ \hline
			Overlap removal      &   remove jet if $\Delta R$(electron, jet) $< 0.2$\\ 
			&  then, remove lepton if $\Delta R$(lepton, jet) $< 0.4$ \\ 
			\hline \hline
			\multicolumn{2}{|c|}{\textbf{Event Selection}} \\
			\hline 
			\multicolumn{2}{|c|}{leading lepton $\pt > 100$~GeV (flavour defines channel)} \\ \hline
			\multicolumn{2}{|c|}{number of objects above 100 GeV $> 2$} \\ \hline
			\multicolumn{2}{|c|}{appropriate $\sumpt$ threshold} \\ \hline
			
		\end{tabular}
	\end{center}
	\caption{Event selection defining the fiducial regions, used in the determination of the reconstruction efficiency $C_{pp\to \ell X}$. All selection requirements are at generator level.}
	\label{tab:truthAnalysisDefinition}
\end{table*}

\begin{table}[htb]
	\begin{center}
		\begin{tabular}{|c||ccc|} \hline
			& \multicolumn{3}{c|}{$\sigma^\mathrm{fid}_{pp\to \ell X}$ $95\%$ CL Upper Limit [fb]}  \\ 
			Min.~$\sumpt$ & \multicolumn{3}{c|}{Observed (Expected)}  \\
			$[$GeV$]$                 & Electron  & Muon & Combined \\
			& Channel & Channel & Channels \\ \hline\hline

			$2000$   &  $	3.04	$	$\left(	2.77	^{	+	0.94	}	_	{	-0.69	}	\right)$	&	$	4.18	$	$\left(	3.44	^{	+	1.16	}	_	{	-0.94	}	\right)$ &  $5.17$ 	$\left( 4.47	^{	+ 1.57	}	_	{ -1.27	}	\right)$ \\    
			$2200$   &  $	1.96	$	$\left(	1.67	^{	+	0.61	}	_	{	-0.43	}	\right)$	&	$	2.62	$	$\left(	2.16	^{	+	0.90	}	_	{	-0.43	}	\right)$ &   $3.22$ 	$\left( 2.67	^{	+ 0.99	}	_	{ -0.74	}	\right)$ \\     
			$2400$   &  $	0.72	$	$\left(	0.80	^{	+	0.33	}	_	{	-0.19	}	\right)$	&	$	2.15	$	$\left(	1.61	^{	+	0.61	}	_	{	-0.41	}	\right)$  &   $1.59$ 	$\left( 1.59	^{	+ 0.60	}	_	{ -0.39	}	\right)$   \\
			$2600$   &  $	0.51	$	$\left(	0.55	^{	+	0.21	}	_	{	-0.14	}	\right)$	&	$	1.15	$	$\left(	1.12	^{	+	0.38	}	_	{	-0.21	}	\right)$   &   $0.89$ 	$\left( 0.99	^{	+ 0.36	}	_	{ -0.26	}	\right)$   \\
			$2800$   &  $	0.32	$	$\left(	0.32	^{	+	0.13	}	_	{	-0.01	}	\right)$	&	$	1.19	$	$\left(	0.92	^{	+	0.33	}	_	{	-0.13	}	\right)$  &   $0.72$ 	$\left( 0.72	^{	+ 0.28	}	_	{ -0.17	}	\right)$   \\
			$3000$   &  $	0.31	$	$\left(	0.31	^{	+	0.01	}	_	{	-0.01	}	\right)$	&	$	0.68	$	$\left(	0.68	^{	+	0.01	}	_	{	-<0.01	}	\right)$ &   $0.43$ 	$\left( 0.44	^{	+ 0.14	}	_	{ -0.01	}	\right)$  \\
			$3200$   &  $	0.31	$	$\left(	0.31	^{	+	0.01	}	_	{	-0.01	}	\right)$	&	$	0.67	$	$\left(0.67	^{	+	0.01	}	_	{	-0.01	}	\right)$  &   $0.43$ 	$\left( 0.43	^{	+ 0.01	}	_	{ -<0.01	}	\right)$  \\

			\hline\hline

		\end{tabular}
		\caption{The observed and expected $95\%$ CL upper limits on the fiducial cross-sections ($\sigma^\mathrm{fid}_{pp\to \ell X}$) for the production of final states with at least three objects passing a $100$~GeV $\pt$ requirement, including at least one isolated lepton, and $\sumpt$ above threshold. The limits are shown for the muon and electron channels separately and combined, using the most conservative reconstruction efficiency. 
The CL$_{\mathrm s}$ method is used to obtain the limits.
		}
		\label{tab:fidexclusion}
	\end{center}
\end{table}

Exclusion contours are obtained in the plane of $\MD$ and $\MTH$ for several benchmark models. In each of the channels, the signal region above $\sumpt=2000$~GeV is divided into multiple slices, with $\sumpt$ thresholds increasing in steps of $200\GeV$. This allows the analysis to be sensitive across a wider range of signal models, and values of \MD and \MTH. 

For each point in the $\MD$-$\MTH$ plane, the $\sumpt$ slice that gives the best expected limit is used.\footnote{The combined exclusions use the same thresholds in the two channels.} The resulting exclusions for the statistical combination of the electron and muon channels, for benchmark black hole models simulated with \charybdis, for two, four and six extra dimensions, are shown in figure~\ref{fig:limits_combined_Charybdis2}. The exclusions tend to be stronger for higher $n$, due to the larger signal cross-sections. They also tend to be stronger for the non-rotating case than for the rotating case, due to a larger number of Hawking emissions (and consequently higher object multiplicity and probability of a leptonic final state). This is most apparent for low values of $\MTH/\MD$, where there are the fewest Hawking emissions (and where the semiclassical production assumptions are least valid). The lower panes of figure~\ref{fig:limits_combined_Charybdis2} show the equivalent exclusions for rotating black holes with low-multiplicity remnant states and with the \charybdis model of production losses. A low multiplicity remnant state weakens the exclusion reach, due to the reduced number of objects and leptonic event fraction, whilst the suppressed cross-section of the latter benchmark model reduces the exclusion by about 0.5 TeV for a given value of $\MD$. 
Exclusion limits for string ball models with six extra dimensions are shown in figure~\ref{fig:limits_combined_stringballs}. Again, the exclusion for the non-rotating case is slightly stronger than for the rotating case.

The equivalent black hole exclusion contours for \blackmax models are given in figure~\ref{fig:limits_combined_BlackMax}. The exclusions show the same general features as the ones obtained with the samples generated by \charybdis. The \blackmax model of production losses does not result in a reduction of the exclusion reach, due to the emission of photons, rather than gravitons. Graviton emission in non-rotating black hole models weakens the exclusion slightly, as a greater number of decay products carry missing, rather than visible, energy.  Figure~\ref{fig:limits_combined_c1} shows limits for a rotating black hole model with lepton conservation imposed, for four extra dimensions. The effect on the excluded region for this model choice is small, showing that the analysis is not sensitive to this model assumption.

Tables \ref{tab:resultsCharybdis} and \ref{tab:resultsBlackmax} summarise the exclusion limits for all models considered in the $n = 6$ case. For \MD = 1.5 TeV, threshold masses below $5.5$--$6.2$~TeV are excluded at 95\% CL, depending on model assumptions. For $\MD = 4$~TeV, threshold masses below $4.8$--$5.7$~TeV are excluded.

\begin{table}[htb]
	\begin{center}
		\begin{tabular}{|c|c||c|c|}\hline
			\multirow{2}{*}{Angular Mom.} & \multirow{2}{*}{Description} & \multicolumn{2}{c|}{Excluded \MTH value [TeV] for:} \\ 
			
			&  & \MD = 1.5 TeV & \MD = 4 TeV\\ \hline

			Non-rotating & Black holes: High multiplicity remnant   & 6.2 & 5.7  \\ 
			Rotating & Black holes: High multiplicity remnant  & 6.0  & 5.4  \\
			Rotating & Black holes: Low multiplicity remnant & 6.0   & 5.2 \\
			Rotating & Production loss model (gravitons)    & 5.5  & 4.8  \\ \hline 
			
			\multicolumn{2}{|c||}{} & \MS = 1.2 TeV & \MS = 2.5 TeV\\ \hline
			Non-rotating & \multirow{2}{*}{String balls}  & 5.7  & 5.1 \\
			Rotating &    & 5.5 & 4.7  \\ 
			
			\hline
		\end{tabular}
		\caption{Limits for $n = 6$ for the \charybdis models detailed in section~\ref{sec:mcsamples}. }
		\label{tab:resultsCharybdis}
	\end{center}
\end{table}

\begin{table}[htb]
	\begin{center}
		\begin{tabular}{|c|c||c|c|}\hline
			\multirow{2}{*}{Angular Mom.} & \multirow{2}{*}{Description}  &\multicolumn{2}{c|}{Excluded \MTH value [TeV] for:} \\ 
			
			&  & \MD = 1.5 TeV & \MD = 4 TeV\\ \hline

 			Non-rotating & Black holes: High multiplicity remnant & 6.2  & 5.6  \\ 
 			Rotating & Black holes: High multiplicity remnant  & 6.1  & 5.6 \\
 			Non-rotating & Black holes with graviton  & 6.2 & 5.6  \\
			Rotating & 10\% Production loss model (photons)  & 6.1  & 5.5 \\ 
			\hline
		\end{tabular}
		\caption{Limits for $n = 6$ for the \blackmax models detailed in section~\ref{sec:mcsamples}. }
		\label{tab:resultsBlackmax}
	\end{center}
\end{table}

\clearpage

%%%%%%%%%%%%%%%%%%%%%%%%%%%%%%%%%%%%%%%%%%%%%%%%%%%%%%%%%%%%%%%%%%%%%%%%%%%%%%
\newpage
\section{Summary}

A search for microscopic black holes and string ball states in the ATLAS detector at the LHC using a total proton--proton integrated luminosity of {\lumi}~fb$^{-1}$ at $\sqrt{s} = 8$ TeV is presented.
The search considers final states with three or more high-$\pt$ objects, at least one of which is required to be an electron or a muon. No deviation from the Standard Model is observed in either the electron or the muon channel. Consequently, limits are set on a wide range of black hole and string ball models. For rotating black holes and six extra dimensions, mass thresholds below $6.0$~TeV ($5.4$~TeV) are excluded for a fundamental Planck scale \MD of $1.5$~TeV ($4.0$~TeV). Upper limits, at $95\%$ CL, are also set on the fiducial cross-sections for new-physics production of high-$\sumpt$ multi-object final states containing a high-$\pt$ (above 100~GeV) lepton within the experimental acceptance. For final states with $\sumpt>3200$~GeV a limit of $0.43$~fb is set considering both channels combined. The limits set by this analysis are the strongest published by the ATLAS Collaboration, and consider a substantially wider range of extra-dimensional models. Together with the limits published by the CMS Collaboration, they represent the strongest current bounds on these models.

\section{Acknowledgements}

We thank CERN for the very successful operation of the LHC, as well as the
support staff from our institutions without whom ATLAS could not be
operated efficiently.

We acknowledge the support of ANPCyT, Argentina; YerPhI, Armenia; ARC,
Australia; BMWF and FWF, Austria; ANAS, Azerbaijan; SSTC, Belarus; CNPq and FAPESP,
Brazil; NSERC, NRC and CFI, Canada; CERN; CONICYT, Chile; CAS, MOST and NSFC,
China; COLCIENCIAS, Colombia; MSMT CR, MPO CR and VSC CR, Czech Republic;
DNRF, DNSRC and Lundbeck Foundation, Denmark; EPLANET, ERC and NSRF, European Union;
IN2P3-CNRS, CEA-DSM/IRFU, France; GNSF, Georgia; BMBF, DFG, HGF, MPG and AvH
Foundation, Germany; GSRT and NSRF, Greece; ISF, MINERVA, GIF, I-CORE and Benoziyo Center,
Israel; INFN, Italy; MEXT and JSPS, Japan; CNRST, Morocco; FOM and NWO,
Netherlands; BRF and RCN, Norway; MNiSW and NCN, Poland; GRICES and FCT, Portugal; MNE/IFA, Romania; MES of Russia and ROSATOM, Russian Federation; JINR; MSTD,
Serbia; MSSR, Slovakia; ARRS and MIZ\v{S}, Slovenia; DST/NRF, South Africa;
MINECO, Spain; SRC and Wallenberg Foundation, Sweden; SER, SNSF and Cantons of
Bern and Geneva, Switzerland; NSC, Taiwan; TAEK, Turkey; STFC, the Royal
Society and Leverhulme Trust, United Kingdom; DOE and NSF, United States of
America.

The crucial computing support from all WLCG partners is acknowledged
gratefully, in particular from CERN and the ATLAS Tier-1 facilities at
TRIUMF (Canada), NDGF (Denmark, Norway, Sweden), CC-IN2P3 (France),
KIT/GridKA (Germany), INFN-CNAF (Italy), NL-T1 (Netherlands), PIC (Spain),
ASGC (Taiwan), RAL (UK) and BNL (USA) and in the Tier-2 facilities
worldwide.

\clearpage

\bibliographystyle{atlasBibStyleWithTitle}
\bibliography{LJPaper2013}
\clearpage
% ATLAS Collaboration author list
% Data extracted on 07-Aug-2014 for paper reference EXOT-2013-05
%\documentclass[11pt]{article}
%\usepackage{a4wide}\begin{document}
\begin{flushleft}
{\Large The ATLAS Collaboration}

\bigskip

G.~Aad$^{\rm 84}$,
B.~Abbott$^{\rm 112}$,
J.~Abdallah$^{\rm 152}$,
S.~Abdel~Khalek$^{\rm 116}$,
O.~Abdinov$^{\rm 11}$,
R.~Aben$^{\rm 106}$,
B.~Abi$^{\rm 113}$,
M.~Abolins$^{\rm 89}$,
O.S.~AbouZeid$^{\rm 159}$,
H.~Abramowicz$^{\rm 154}$
% ...
\bigskip
\\
% ...
$^{11}$ Institute of Physics, Azerbaijan Academy of Sciences, Baku, Azerbaijan\\
% ...
$^{84}$ CPPM, Aix-Marseille Universit{\'e} and CNRS/IN2P3, Marseille, France\\
% ...
$^{89}$ Department of Physics and Astronomy, Michigan State University, East Lansing MI, United States of America\\
% ...
$^{106}$ Nikhef National Institute for Subatomic Physics and University of Amsterdam, Amsterdam, Netherlands\\
% ...
$^{112}$ Homer L. Dodge Department of Physics and Astronomy, University of Oklahoma, Norman OK, United States of America\\
$^{113}$ Department of Physics, Oklahoma State University, Stillwater OK, United States of America\\
% ...
$^{116}$ LAL, Universit{\'e} Paris-Sud and CNRS/IN2P3, Orsay, France\\
% ...
$^{152}$ Institute of Physics, Academia Sinica, Taipei, Taiwan\\
% ...
$^{154}$ Raymond and Beverly Sackler School of Physics and Astronomy, Tel Aviv University, Tel Aviv, Israel\\
% ...
$^{159}$ Department of Physics, University of Toronto, Toronto ON, Canada\\
% ...
\end{flushleft}

%\end{document}
% Created with ./xml2latex.py

\end{document}